\documentclass[prl,preprint,endfloats]{revtex4}

\usepackage[dvipdfmx]{graphicx}
\usepackage{bm}
\usepackage{pifont}
\usepackage{amsmath}
\usepackage{color}
\usepackage{comment}
\usepackage{transparent}

\begin{document}

\title{Multi-step topological transitions among meron and skyrmion crystals in a centrosymmetric magnet}

\author{H. Yoshimochi$^{1}$, R. Takagi$^{1,2,3}$, J. Ju$^{4}$, N. D. Khanh$^{5}$, H. Saito$^{4}$, H. Sagayama$^{6}$, H. Nakao$^{6}$, S. Itoh$^{6,7}$, Y. Tokura$^{1, 5, 8}$, T. Arima$^{5, 9}$, S. Hayami$^{10}$, T. Nakajima$^{4, 5}$, S. Seki$^{1, 2, 3}$}

\affiliation{$^1$ Department of Applied Physics, University of Tokyo, Tokyo 113-8656, Japan, \\ 
$^2$ Institute of Engineering Innovation, University of Tokyo, Tokyo 113-0032, Japan, \\ 
$^3$ PRESTO, Japan Science and Technology Agency (JST), Kawaguchi 332-0012, Japan, \\ 
$^4$ Institute for Solid State Physics, University of Tokyo, Kashiwa 277-0882, Japan,\\
$^5$ RIKEN Center for Emergent Matter Science (CEMS), Wako 351-0198, Japan,\\
$^6$ Institute of Materials Structure Science, High Energy Accelerator Research Organization, Tsukuba 319-1195, Japan,\\
$^7$ Materials and Life Science Division, J-PARC Center, Tokai, Ibaraki 319-1195, Japan, \\
$^8$ Tokyo College, University of Tokyo, Tokyo 113-8656, Japan,\\
$^9$ Department of Advanced Materials Science, University of Tokyo, Kashiwa 277-8561, Japan,\\
$^{10}$ Graduate School of Science, Hokkaido University, Sapporo 060-0810, Japan}

\begin{abstract}

{\bf Topological swirling spin textures, such as skyrmions and merons, have recently attracted much attention as a unique building block for high-density magnetic information devices. The controlled transformation among different types of such quasi-particles is an important challenge, while it was previously achieved only in a few non-centrosymmetric systems characterized by Dzyaloshinskii-Moriya interaction. Here, we report an experimental discovery of multi-step topological transitions among a variety of meron and skyrmion crystal states in a centrosymmetric magnet GdRu$_2$Ge$_2$. By performing the detailed magnetic structure analysis based on resonant X-ray and neutron scattering experiments as well as electron transport measurements, we have found that this compound hosts periodic lattice of elliptic skyrmions, meron/anti-meron pairs, and circular skyrmions as a function of external magnetic field. The diameter of these objects is as small as 2.7 nm, which is almost two orders of magnitude smaller than typical non-centrosymmetric magnets. Such an intricate manner of topological magnetic transitions are well reproduced by a theoretical model considering the competition between RKKY interactions at inequivalent wave vectors. The present findings demonstrate that even a simple centrosymmetric magnet with competing interactions can be a promising material platform to realize a richer variety of nanometric magnetic quasi-particles with distinctive symmetry and topology, whose stability may be tunable by various external stimuli.}

\end{abstract}

\maketitle

Topological swirling textures, such as skyrmions, merons, and hopfions, have recently been explored in various condensed matter systems with broken space-inversion symmetry (such as magnets\cite{SkXReviewFertTwo, SkXReviewTokura, OlegReview, Meron1, Meron2}, ferroelectrics\cite{PolarSkyrmion}, and chiral liquid crystals\cite{LQ_Meron, LQ_3D, LQ_3D2}). These objects and their molecule-like clusters\cite{SkBag, Bimeron} often appear in the form of isolated particles and periodic lattice. They can be a source of rich emergent phenomena, and better understanding of their formation mechanism and unique response to external stimuli is one of the central issues in material science.

Recently, skyrmions in magnetic materials (Fig. 1b) have attracted attention as a potential candidate of high-density information carrier, because of their small size and unique electric controllability\cite{SkXReviewFertTwo, SkXReviewTokura, OlegReview, SkXReviewKanazawa, SkXTheoryFirst, MnSi, TEMFeCoSi}. Magnetic skyrmions are characterized by the topological charge defined as
\begin{equation}
   N_{{\rm sk}} = \frac{1}{4\pi} \int {\bf n} \cdot \left(\frac{\partial {\bf n}}{\partial x} \times \frac{\partial {\bf n}}{\partial y}\right) {\rm d}x {\rm d}y,
\end{equation}
which represents how many times the spin directions wrap a sphere. ${\bf n}({\bf r})={\bf m}({\bf r})/|{\bf m}({\bf r})|$ is the unit vector along the local magnetic moment ${\bf m}({\bf r})$, and the integral is taken over a selected two-dimensional area. In general, skyrmion, anti-skyrmion, meron, and anti-meron represent the spin texture ${\bf m}({\bf r})$ characterized by $N_{\rm sk} = -1, +1, -1/2$ and $+1/2$, respectively\cite{SkXReviewTokura, Meron1, Meron2}. Figures 1b and c schematically illustrate skyrmion and anti-skyrmion spin textures, accompanied with vortex and anti-vortex type arrangement of in-plane spin components, respectively. In both cases, ${\bf m}({\bf r})$ at the core (edge) region is anti-parallel (parallel) to the out-of-plane external magnetic field ${\bf B}$. On the other hand, (anti)meron corresponds to half (anti)skyrmion, and they often appear in the form of molecule-like pair\cite{Bimeron, TypeIIBubble}. The examples of (anti)meron spin textures are shown in Figs. 1d-g. (See Supplementary Note V for the detailed description of merons and anti-merons.)

Magnetic skyrmions were originally discovered in a series of non-centrosymmetric compounds, where Dzyaloshinskii-Moriya (DM) interaction plays a crucial role in the skyrmion formation\cite{SkXReviewKanazawa, MnSi, TEMFeCoSi}. In the zero-field state, these compounds commonly host helical spin order characterized by a single magnetic modulation vector ${\bf Q}$. The application of moderate amplitude of external magnetic field $B$ induces a skyrmion lattice state, which can be approximately described by the superposition of multiple spin helices modulated along different orientations (i.e. multi-{\bf Q} state)\cite{MnSi}. Their spin swirling manner is governed by the symmetry of DM interaction\cite{SkXReviewKanazawa}, and skyrmion (Figs. 1b) and anti-skyrmion (Figs. 1c) spin textures are usually stabilized in the system with chiral\cite{MnSi, TEMFeCoSi, Cu2OSeO3_Seki, CoZnMn_First} and $D_{2d}$/$S_4$\cite{Heusler, S4Karube} crystal symmetries, respectively. In this mechanism, a typical skyrmion diameter is in the order of several tens to hundreds nanometer.

On the other hand, recent theoretical studies have suggested that skyrmions can be stabilized even without inversion symmetry breaking, owing to different microscopic mechanisms\cite{HayamiModel1, HayamiModel2, BatistaModel, Frustration1, Frustration2, BatistaModel_Frustration, Blugel_Nesting}. In particular, extremely small diameter (less than 3 nm) of skyrmions have been discovered in a few centrosymmetric rare-earth intermetallic compounds, such as hexagonal Gd$_2$PdSi$_3$\cite{Gd2PdSi3} and tetragonal GdRu$_2$Si$_2$\cite{GdRu2Si2}. For the latter centrosymmetric systems, it has been proposed that itinerant-electron-mediated interactions play an important role, while their detailed skyrmion formation mechanism is still in controversy\cite{Gd2PdSi3, GdRu2Si2, GdKagome, EuPtSi, GdRu2Si2_STM, GdRu2Si2_Full, EuAl4, TbMn6Sn6}.

Here, one of the key challenges is the identification of a general strategy to realize a richer variety of magnetic quasi-particles in such systems. So far, most of the reported materials host only a circular skyrmion phase in equilibrium\cite{SkXReviewKanazawa}, and other types of magnetic quasi-particles have rarely been explored experimentally. Only recently, $B$-induced transition between the square meron lattice and hexagonal skyrmion lattice states was discovered in a non-centrosymmetric DM magnet Co$_8$Zn$_9$Mn$_3$\cite{Meron1}, while it is rather exceptional. The controlled transformation among different types of solitonic spin textures may lead to the multiple-valued memory function\cite{OlegReview}, and further search of novel materials and mechanisms to realize a wider variety of exotic magnetic quasi-particles is highly anticipated.

In this study, we report the experimental discovery of multi-step topological transitions among elliptic skyrmion, meron/anti-meron pair, and circular skyrmion phases in a centrosymmetric magnet GdRu$_2$Ge$_2$. Our detailed theoretical analysis has revealed that such an intricate manner of magnetic transitions are well reproduced by considering the competition between RKKY interactions at inequivalent wave vectors ${\bf Q}_{\rm A} = (q, 0, 0)$ and ${\bf Q}_{\rm B} = (q/2, q/2, 0)$. The present findings demonstrate that even a simple centrosymmetric compound with competing interactions can be a promising material platform to realize a richer variety of nanometric magnetic quasi-particles with distinctive symmetry and topology, whose stability may be tunable by various external stimuli.

Our target material GdRu$_2$Ge$_2$ is characterized by ThCr$_2$Si$_2$-type crystal structure of centrosymmetric tetragonal space group $I4/mmm$, as shown in Fig. 1a. It consists of alternate stacking of square lattice Gd layers and Ru$_2$Ge$_2$ layers, and the magnetism is governed by Gd$^{3+}$ ($S$ = 7/2, $L$ = 0) ions with an approximately isotropic magnetic moment. According to a previous magnetization measurement\cite{GdRu2Ge2_PhaseDiagram}, this compound shows several magnetic transitions in $B \parallel [001]$, while the detailed magnetic structure in each phase and associated electron transport properties yet remain to be clarified.

Figures 1j and k show the magnetic-field dependence of magnetization $M$, longitudinal resistivity $\rho_{xx}$ and Hall resistivity $\rho_{yx}$ measured at 6 K for $B \parallel [001]$. The magnetization profile shows three distinctive intermediate steps at 1.0 T, 1.2 T and 1.35 T, as well as several additional kinks, before reaching the saturated ferromagnetic (FM) state with $M \approx $ 7 $\mu_{\rm B}$/Gd$^{3+}$ above 4.5 T. They represent the successive metamagnetic phase transitions among Phases I, II, III, IV, V, VI and the FM state\cite{GdRu2Ge2_PhaseDiagram}. Corresponding anomalies are also observed in the $\rho_{xx}$ and $\rho_{yx}$ profiles. In particular, peak-like enhancements of $\rho_{yx}$ can be identified in Phases II and IV. In general, Hall resistivity $\rho_{yx}$ can be described\cite{SkXReviewTokura, THE, RMP_AHE} as
\begin{equation}
\rho_{yx} = \rho_{yx}^{{\rm N}}+\rho_{yx}^{{\rm A}}+\rho_{yx}^{{\rm T}} = R_0 B + R_s M + P R_0 B_{\rm em},
\end{equation}
where $\rho_{yx}^{{\rm N}}$ and $\rho_{yx}^{{\rm A}}$ are the normal Hall term proportional to $B$ and anomalous Hall term proportional to $M$, respectively ($R_0$ and $R_s$ are the coefficients for the respective terms). The third term $\rho_{yx}^{{\rm T}}$ represents the topological Hall effect, which is allowed to appear for nontrivial spin textures with $N_{\rm sk} \neq 0$. When conduction electrons pass through such a topological spin texture, they are expected to obtain an additional quantum-mechanical Berry phase and feel a fictitious emergent magnetic field $B_{\rm em}$ in proportion to the topological charge density. It leads to the appearance of topological Hall term $\rho_{yx}^{{\rm T}}$ proportional to $B_{\rm em}$ ($P$ represents the spin polarization ratio of conduction electrons)\cite{SkXReviewTokura, THE, THE_PRB2}. As detailed in Supplementary Note X, our analysis suggests that the observed Hall profile cannot be simply explained by $\rho_{yx}^{{\rm N}}$ and $\rho_{yx}^{{\rm A}}$, and the peak-like enhancements of $\rho_{yx}$ in Phases II and IV probably originate from $\rho_{yx}^{{\rm T}}$. It implies the appearance of topological spin textures with $N_{\rm sk} \neq 0$ in these phases. We have also performed similar measurements at various temperatures. The result is summarized as a $B$-$T$ magnetic phase diagram for $B \parallel [001]$ shown in Figs. 1h and i. Here, the magnetic phase boundaries are determined based on the magnetization data, and the background color indicates the value of Hall resistivity $\rho_{yx}$. It evidences the strong correlation between the magnetism and electrical transport properties, and also confirms the clear enhancements of $\rho_{yx}$ in Phases II and IV.

To investigate the magnetic structure in each phase, we first performed neutron scattering experiments at 6 K for various amplitude of $B \parallel [001]$, and measured intensity profiles around reciprocal lattice points $(-1,1,0)$ and $(0,-1,-1)$ as detailed in Supplementary Fig. 8 and Supplementary Note VII. We found that all magnetic phases except the FM phase are characterized by an in-plane magnetic modulation vector ${\bf Q} \sim (q,0,0)$ or $(0,q,0)$. In Phases II, III, and IV, additional magnetic peaks approximately indexed as ${\bf Q} \sim (q/2,q/2,0)$ or $(-q/2,q/2,0)$ are also observed. In general, the multi-${\bf Q}$ magnetic order characterized by multiple fundamental modulation vectors ${\bf Q}_\nu$ and ${\bf Q}_{\nu '}$ is allowed to induce the higher order reflections ${\bf Q}_\nu \pm {\bf Q}_{\nu '}$\cite{GdRu2Si2, MnSi_LongRange}, and the neutron scattering pattern in Supplementary Fig. 8 implies the multi-${\bf Q}$ character of Phases II, III and IV.

Next, these magnetic satellite peaks are investigated in more detail by means of resonant X-ray (RXS) scattering. The energy of the incident X-ray beam is tuned to the Gd-$L_2$ absorption edge. The positions of magnetic reflections for Phases I, II, III, IV and V observed in the RXS measurement are schematically illustrated in Figs. 2o-s. We focused on satellite reflections near the reciprocal lattice points of $(4, 2, 0)$ and $(4, 0, 0)$, and performed two types of line scans A and B corresponding to the green and blue arrows in Fig. 2o, respectively, as discussed below. 

The ($4+\delta$, 2, 0) line profile (i.e. line scan A) measured at $B=0$ (Phase I) is shown in Fig. 2e. A sharp reflection peak suggests that Phase I is characterized by a magnetic modulation vector ${\bf Q}_{\rm A} = (q, 0, 0)$ with $q \sim 0.213$. Measurements along the same line were also performed for various amplitudes of $B \parallel [001]$, and the obtained line profiles for Phases II, III, IV and V are indicated in Figs. 2f-i. On the basis of these data, the magnetic-field dependence of the wave number $q$ and integrated intensity of $(4+q, 2, 0)$ magnetic reflection, as well as magnetization $M$, are plotted in Figs. 2b-d. We have found that the magnetic phase transitions characterized by step-like magnetization anomalies are accompanied by abrupt changes in $q$-value and scattering intensity. Notably, the magnetic reflection on the line A splits into two peaks with distinctive $q$-values $q_1$ and $q_2$ ($q_1 < q_2$) in Phases II and III (Figs. 2f and g).

To investigate the possible multi-${\bf Q}$ character in each phase, the (4-$\tau$, -$q_0 + \tau$, 0) line profiles are further measured (i.e. line scan B) as summarized in Figs. 2j-n. We define $q_0 = (q_1 + q_2)/2$ based on the wave numbers $q_1$ and $q_2$ identified from the line scan A, where $q=q_1=q_2$ holds for Phases I, IV and V. Our measurements reveal that magnetic reflections on the line B are observable only in Phases II, III and IV, in accord with the neutron data in Supplementary Fig. 8. In Phase IV, the peak position on the line B can be exactly indexed as ${\bf Q}_{\rm B} = (q/2, q/2, 0)$. By considering the symmetrically equivalent wave vector ${\bf Q}_{\rm B'} = (-q/2, q/2, 0)$, the relation ${\bf Q}_{\rm A} = {\bf Q}_{\rm B} - {\bf Q}_{\rm B'}$ is satisfied (Fig. 2r). It indicates that Phase IV hosts multi-${\bf Q}$ magnetic order. As detailed in Supplementary Note II, similar relations are also satisfied in Phases II and III (Figs. 2p and q), where the peak positions on the line B are indexed as ${\bf Q}_{\rm B} = (q_1/2, q_2/2, 0)$ or $(q_2/2, q_1/2, 0)$. It also demonstrates the multi-${\bf Q}$ character of these phases. (Note that the appearance of higher order reflection is not the necessary condition of multi-${\bf Q}$ state, and Phase V is also assigned as a multi-${\bf Q}$ state as discussed in Supplementary Note I.)

When the magnetic structure breaks the symmetry of the original crystal structure, there should appear multiple equivalent magnetic domains that are converted into each other by the broken symmetry elements. In case of Phase I (Fig. 2o), the single-${\bf Q}$ magnetic order breaks the four-fold symmetry, and two magnetic domains with ${\bf Q}_{\rm A} = (q, 0, 0)$ and $(0, q, 0)$ can coexist. In a similar manner, the four-fold symmetry is broken in Phases II and III. It leads to the emergence of two magnetic domains $\alpha$ and $\beta$ related to each other by four-fold rotation, whose magnetic satellite positions are denoted by black and white circles in Figs. 2p and q, respectively (See Supplementary Note II and Supplementary Fig. 2 for the detail). It explains the observed appearance of two distinctive magnetic reflections for the line scan A in Phases II and III (Figs. 2f and g), where two peaks reflect the contribution from the domains $\alpha$ and $\beta$. In Phase IV (Fig. 2r), the four-fold symmetry is recovered and the consideration of such rotational domains is not necessary.

Next, to identify the detailed spin orientations in each phase, we perform the polarization analysis of the scattered X-ray beam. The measurement configuration is illustrated in Fig. 2a, in which the propagation vectors of the incident and scattered X-ray beams (${\bf k}_i$ and ${\bf k}_f$, respectively) are always confined within the (001) plane. Here, the incident 
 X-ray beam is linearly polarized parallel to the scattering plane ($\pi$-polarized). The scattered beam includes two polarization components parallel ($\pi'$) and perpendicular ($\sigma'$) to the scattering plane, and their intensities ($I_{\pi-\pi'}$ and $I_{\pi-\sigma'}$) are measured separately. When the magnetic structure ${\bf m} ({\bf r})$ is composed of modulated spin component ($\tilde{{\bf m}} ({\bf Q}) {\rm exp}(i{\bf Q} \cdot {\bf r})+$ c.c.) with the wave vector ${\bf Q}$, the corresponding magnetic scattering intensity is described\cite{RXS_Rule} as
\begin{equation}
\label{scatInt}
I \propto |({\bf e}_i \times {\bf e}_f) \cdot \tilde{{\bf m}}({\bf Q})|^2
\end{equation}
with ${\bf e}_i$ and ${\bf e}_f$ representing the polarization vectors of the incident and scattered beams, respectively. Here, $\tilde{{\bf m}}({\bf Q})$ is a complex vector and c.c. represents the complex conjugate. In general, $\tilde{{\bf m}}$(${\bf Q}$) can be decomposed as
\begin{equation}
\tilde{{\bf m}}({\bf Q})= \tilde{m}_{c}({\bf Q}) {\bf e}_c + \tilde{m}_{Q}({\bf Q}) {\bf e}_Q + \tilde{m}_{c \times Q}({\bf Q}) {\bf e}_{c \times Q},
\end{equation}
where ${\bf e}_c$, ${\bf e}_Q$, and ${\bf e}_{c \times Q}$ ($\tilde{m}_c$, $\tilde{m}_Q$, and $\tilde{m}_{c \times Q}$) are the unit vectors (amplitudes of modulated spin components) parallel to the [001] axis, ${\bf Q}$-vector, and the axis perpendicular to both of them, respectively. In the present setup with the (001) scattering plane (Fig. 2a), $I_{\pi-\pi'}$ always represents the $\tilde{m}_{c}$ component, and $I_{\pi-\sigma'}$ reflects the component of $\tilde{{\bf m}}({\bf Q})$ parallel to ${\bf k}_i$.

First, we investigate Phase IV with four-fold symmetric multi-${\bf Q}$ character. For this purpose, the magnetic satellites at the $(4, 2, 0) \pm {\bf Q}$ position are studied, with the measurement geometry shown in Fig. 3a. In this configuration, ${\bf k}_i$ is almost parallel to the [100] axis, and $I_{\pi-\sigma'}$ mainly reflects the [100] component of $\tilde{{\bf m}}({\bf Q})$. Figures 3b and d show the ($4+\delta$, 2, 0) line profile measured in Phase IV, where the presence of $I_{\pi-\pi'}$ (absence of $I_{\pi-\sigma'}$) suggests the presence of $\tilde{m}_{c}$ (the absence of $\tilde{m}_{Q}$) for ${\bf Q}_{\rm A} = (q, 0, 0)$. The corresponding (4, $2+\delta$, 0) line profile is also shown in Figs. 3c and e, where the presence of $I_{\pi-\pi'}$ and $I_{\pi-\sigma'}$ indicate the presence of $\tilde{m}_{c}$ and $\tilde{m}_{c\times Q}$ for ${\bf Q}_{\rm A'} = (0, q, 0)$. These results suggest that ${\bf Q}_{\rm A}$ and ${\bf Q}_{\rm A'}$ are characterized by the screw-type spin modulation, as shown in Fig. 3f, where their neighboring spin components rotate within a plane normal to each wave vector. 

In a similar manner, the magnetic satellites at the $(4, 0, 0) \pm {\bf Q}$ positions have also been investigated, with the measurement geometry shown in Fig. 3m. In this case, ${\bf k}_i$ is almost parallel to the $[1\bar{1}0]$ axis, and $I_{\pi-\sigma'}$ mainly reflects the $[1\bar{1}0]$ component of $\tilde{{\bf m}}({\bf Q})$. Figures 3n and p show the $(4-\tau, q_0-\tau, 0)$ line profile measured in Phase IV, where the presence of $I_{\pi-\pi'}$ (absence of $I_{\pi-\sigma'}$) suggests the presence of $\tilde{m}_{c}$ (the absence of $\tilde{m}_{Q}$) for ${\bf Q}_{\rm B'} = (-q/2, q/2, 0)$. The corresponding $(4-\tau, -q_0+\tau, 0)$ line profile is also shown in Figs. 3o and q, where the presence of $I_{\pi-\pi'}$ and $I_{\pi-\sigma'}$ indicates the presence of $\tilde{m}_{c}$ and $\tilde{m}_{c\times Q}$ for ${\bf Q}_{\rm B} = (q/2, q/2, 0)$. These results demonstrate that ${\bf Q}_{\rm B}$ and ${\bf Q}_{\rm B'}$ are also characterized by the screw-type spin modulation, as shown in Fig. 3r. To summarize, Phase IV can be approximately described as the superposition of four distinctive screw-type spin modulations characterized by ${\bf Q}_{\rm A}$, ${\bf Q}_{\rm A'}$, ${\bf Q}_{\rm B}$ and ${\bf Q}_{\rm B'}$ (Figs. 3f and r).

We have also performed the same line scans for Phases III and II with the anisotropic multi-${\bf Q}$ character, as summarized in Figs. 3g-l and 3s-x. In this case, the four-fold symmetry breaking leads to the appearance of magnetic domains $\alpha$ and $\beta$ (Figs. 3b, c, n and o). Domain $\alpha$ is characterized by  ${\bf Q}_{\rm A} = (q_1, 0, 0)$, ${\bf Q}_{\rm A'} = (0, q_2, 0)$, ${\bf Q}_{\rm B} = (q_1/2, q_2/2, 0)$ and ${\bf Q}_{\rm B'} = (-q_1/2, q_2/2, 0)$, while domain $\beta$ is characterized by ${\bf Q}_{\rm A} = (0, -q_1, 0)$, ${\bf Q}_{\rm A'} = (q_2, 0, 0)$, ${\bf Q}_{\rm B} = (q_2/2, -q_1/2, 0)$ and ${\bf Q}_{\rm B'} = (q_2/2, q_1/2, 0)$. As a result, each line scan profile is characterized by two magnetic satellite peaks reflecting the contributions from the domains $\alpha$ and $\beta$ (Note that the broadness of magnetic reflection peaks in Figs. 3q,t,w and Figs. 2j-n is due to the limited wave number resolution along this scan direction). In Phase II, the selection rule of magnetic scattering (i.e. the presence/absence of $I_{\pi-\pi'}$ and $I_{\pi-\sigma'}$ for each line scan) is exactly the same as the Phase IV, which suggests that Phase II is also characterized by the superposition of four distinctive screw-type spin modulations characterized by ${\bf Q}_{\rm A}$, ${\bf Q}_{\rm A'}$, ${\bf Q}_{\rm B}$ and ${\bf Q}_{\rm B'}$ (Figs. 3l and x). For Phase III, though the selection rule of magnetic scattering is almost the same as Phase IV, the appearance of $I_{\pi-\sigma'}$ in Fig. 3g (at $\delta = q_1 \sim 0.219$) and Fig. 3s suggests the existence of additional small $\tilde{m}_{Q}$ spin component for ${\bf Q}_{\rm A}$, ${\bf Q}_{\rm B}$ and ${\bf Q}_{\rm B'}$. For these wave vectors, screw-type spin spiral plane is slightly tilted toward the ${\bf Q}$ directions, as shown in Figs. 3i and u. On the basis of the observed $I_{\pi-\pi'}$ and $I_{\pi-\sigma'}$ values for each magnetic reflection, the relative amplitudes of $\tilde{m}_{c}$, $\tilde{m}_{Q}$ and $\tilde{m}_{c \times Q}$ are deduced for four distinctive wave vectors in Phases II, III and IV as summarized in Figs. 3f, i, l, r, u and x (See Supplementary Note I and III for the detail, which also includes the results for Phases I and V). In Supplementary Note III and IV, we have further performed a similar analysis for higher order wave vectors such as  ${\bf Q}_{\rm C} = 2{\bf Q}_{\rm B}$ and ${\bf Q}_{\rm C'}=2{\bf Q}_{\rm B'}$.

On the basis of the experimentally deduced $\tilde{m}_{c}$, $\tilde{m}_{Q}$ and $\tilde{m}_{c \times Q}$ for ${\bf Q}_{\rm A}$, ${\bf Q}_{\rm A'}$, ${\bf Q}_{\rm B}$, ${\bf Q}_{\rm B'}$, ${\bf Q}_{\rm C}$ and ${\bf Q}_{\rm C'}$ in each phase (Supplementary Table 1), we reconstruct the corresponding real-space spin texture ${\bf m}({\bf r})$ described as
\begin{equation}
{\bf m(r)} ={\bf e}_c m_c^0 + \sum_{\alpha, \nu} {\bf e}_\alpha \tilde{m}_{\alpha}({{\bf Q}_\nu}) {\rm sin}({\bf Q}_\nu \cdot {\bf r} + \theta_{\alpha}^{{\bf Q}_\nu}).
\label{ReconstructEq}
\end{equation}
Here, $\alpha$ represents the directions $c$, $Q$ or $c\times Q$. $m_c^0$ is the uniform magnetization component in the $B \parallel [001]$ direction, and its value is estimated from the experimental $M$-$B$ profile. While the phase $\theta_{\alpha}^{{\bf Q}_\nu}$ cannot be directly determined from the X-ray scattering experiments, the localized character of Gd$^{3+}$ magnetic moment requires the spatially uniform $|{\bf m}({\bf r})|$ distribution. By considering this constraint, appropriate $\theta_{\alpha}^{{\bf Q}_\nu}$ values can be uniquely identified. For this purpose, we exhaustively investigate various combinations of $\theta_{\alpha}^{{\bf Q}_\nu}$, and deduce the relative phases that provide the most uniform $|{\bf m}({\bf r})|$ distribution (See Supplementary Note III for the detail). 

The resultant real-space spin texture ${\bf m} ({\bf r})$ for Phases I, II, III, IV and V reconstructed based on Eq. (\ref{ReconstructEq}), as well as their schematic illustrations, are summarized in Figs. 4a-e and Figs. 4f-j. The corresponding $|{\bf m} ({\bf r})|$ profiles are shown in Figs. 4k-m, which satisfies almost uniform $|{\bf m}({\bf r})|$ distribution. At $B=0$, the single-${\bf Q}$ screw spin texture is realized in Phase I (Figs. 4a and f). By applying $B \parallel [001]$, helical stripes are pinched off, and turn into the lattice of elliptic skyrmion ($N_{\rm sk}=-1$) with vortex-like arrangement of in-plane spin components as shown in Figs. 4b and g (Phase II). Each elliptic skyrmion transforms into a meron/anti-meron pair ($N_{\rm sk} = 0$) in Phase III (consisting of the combination of Figs. 1d and g), where the in-plane spin component can be described as the combination of a half vortex and a half anti-vortex (Figs. 4c and h). In Phase IV, the pair further turns into a circular skyrmion ($N_{\rm sk}=-1$), as shown in Figs. 4d and i. Finally, the sign of out-of-plane spin component ($m_c$) at the skyrmion core is reversed, and the circular vortex ($N_{\rm sk} = 0$) is realized in Phase V (Figs. 4e and j). In this process, the region with negative $m_c$ (antiparallel to $B$) gradually shrinks as a function of $B$, and the associated Zeeman energy gain can be considered as the main driving force for these magnetic transitions.

Here, Phases II and IV represent the skyrmion lattice states with non-zero net topological charge and skyrmion lattice constant $\sim 2.7$ nm, which is consistent with the observed peak-like enhancement of $\rho_{yx}$ in Figs. 1j and k associated with the topological Hall effect (See Supplementary Note X). In Phase III, the meron and anti-meron form a molecule-like pair, whose core region is characterized by negative $m_c$ component (See Supplementary Note V for the detail). When we consider the skyrmion as meron-meron pair consisting of the combination of Figs. 1d and f, the incremental and decremental change of topological number $\Delta N_{\rm sk} = \pm 1$ for the transitions Phases II $\rightarrow$ III $\rightarrow$ IV can be interpreted as the step-by-step local transformation between the meron ($N_{\rm sk}=-1/2$) and anti-meron ($N_{\rm sk}=+1/2$) spin textures (Figs. 4b-d and g-i). These magnetic phases with distinctive $N_{\rm sk}$ values are separated by sizable energy barrier, as evidenced by the appearance of clear hysteresis in Fig. 1k. Note that the vortex-lattice state in Phase V (Figs. 4e and j) is also characterized by local fractional topological charge, while its spin texture cannot wrap a half of unit sphere due to the uniform out-of-plane magnetization component. Therefore, Phase V cannot be considered as a genuine meron/anti-meron lattice state.

Since GdRu$_2$Ge$_2$ is characterized by the centrosymmetric crystal structure, the observed multi-step topological transitions cannot be explained in terms of DM interaction, and some different microscopic mechanism must be considered. Following Refs. \cite{HayamiModel1, HayamiModel2, Hayami_JPSJ}, we assume the itinerant-electron-mediated interactions on two-dimensional square lattice, and perform the simulated annealing based on the effective magnetic Hamiltonian derived from the Kondo lattice model given by
\begin{equation}
\mathcal{H} = -2J \sum_{\nu, \alpha, \beta} \Gamma^{\alpha \beta}_{{\bf Q}_\nu} {\tilde m}_{\alpha}({{\bf Q}_\nu}) {\tilde m}_{\beta}({{\bf Q}_\nu}) - \sum_i {\bf B} \cdot {\bf m} ({\bf r}_i),
\label{Hamiltonian}
\end{equation}
with $\alpha,\beta = [100], [010], [001]$ and ${\bf Q}_{\nu}= {\bf Q}_{\rm A}, {\bf Q}_{\rm A'}, {\bf Q}_{\rm B}, {\bf Q}_{\rm B'}$. $\tilde{{\bf m}} ({\bf Q}_\nu)$ is the Fourier transform of the real-space distribution of classical localized spin ${\bf m} ({\bf r}_i)$, whose amplitude is fixed at $|{\bf m} ({\bf r}_i)|=1$. Here, the first and second terms represent RKKY interaction and Zeeman coupling, respectively. We suppose the situation as shown in Fig. 5a, where the bare-susceptibility $\chi ({\bf Q})$ of itinerant electron shows the maxima at ${\bf Q}_{\rm A} =(q, 0)$ and ${\bf Q}_{\rm A'} =(0,q)$ with $q = 2\pi/5$, as well as relatively large values at ${\bf Q}_{\rm B} =(q/2, q/2)$ and ${\bf Q}_{\rm B'} =(-q/2, q/2)$ satisfying the relation ${\bf Q}_{\rm A} = {\bf Q}_{\rm B} - {\bf Q}_{\rm B'}$ and ${\bf Q}_{\rm A'} = {\bf Q}_{\rm B} + {\bf Q}_{\rm B'}$. Such $\chi({\bf Q})$ distribution can be naturally realized, for example, by considering the nesting of the Fermi surfaces\cite{HayamiModel1, Hayami_JPSJ}. For these ordering vectors, we adjust the value of interaction tensors $\Gamma^{\alpha \beta}_{{\bf Q}_\nu}$ to satisfy the tetragonal lattice symmetry, weak easy-axis anisotropy\cite{GdRu2Ge2_PhaseDiagram}, and the relation $\chi({\bf Q}_{\rm A}) = \chi({\bf Q}_{\rm A'}) > \chi({\bf Q}_{\rm B}) = \chi({\bf Q}_{\rm B'})$ (See Methods section and Supplementary Note VIII for the detail). 

Figure 5b indicates the magnetic field dependence of magnetization $M$ and associated scalar spin chirality $N'_{\rm sk}$ (which becomes non-zero for $N_{\rm sk} \neq 0$ as discussed in the Methods section) for $B \parallel [001]$ calculated based on Eq. (\ref{Hamiltonian}). It predicts successive magnetic phase transitions (Phase I $\rightarrow$ II $\rightarrow$ III $\rightarrow$ IV $\rightarrow$ V $\rightarrow$ FM), where Phases II and IV are characterized by non-zero scalar spin chiralities. Theoretically obtained spin texture ${\bf m} ({\bf r})$ for each magnetic phase is summarized in Figs. 5c-g, which well reproduces the experimentally deduced ones in Figs. 4a-e. Figures 5h-l indicate the simulated reciprocal space distribution of $|\tilde{{\bf m}}({\bf Q})|^2$, scaling with the scattering intensity expected at each ${\bf Q}$ position in the RXS experiments. Here, magnetic peaks at ${\bf Q}_{\rm B} = (q/2, q/2)$ and ${\bf Q}_{\rm B'} = (-q/2, q/2)$ are observable only in Phases II, III and IV, in accord with the experimental RXS patterns in Figs. 2o-s. In Phases II and III, the four-fold symmetry is broken, and ${\bf Q}_{\rm A} = (q, 0)$ is characterized by a much larger value of $|\tilde{{\bf m}}({\bf Q})|^2$ than ${\bf Q}_{\rm A'} = (0, q)$. It well explains the different scattering intensity between the first and second peaks (corresponding to ${\bf Q}_{\rm A}$ of domain $\alpha$ and ${\bf Q}_{\rm A'}$ of domain $\beta$, respectively) in Figs. 2f and g. The overall good agreement between the theoretical and experimental results supports the validity of our magnetic structure analysis in Fig. 4, and suggests that the magnetism in GdRu$_2$Ge$_2$ is well captured by Eq. (\ref{Hamiltonian}).

Note that earlier theoretical works for the square lattice system\cite{HayamiModel1, HayamiModel2} assumed RKKY interaction at ${\bf Q}_{\rm A}$ and ${\bf Q}_{\rm A'}$ only, and did not consider the ${\bf Q}_{\rm B}$ or ${\bf Q}_{\rm B'}$ contribution. In that case, the large amplitude of four-spin interaction ($\propto \sum_\nu (\sum_{\alpha, \beta} \Gamma^{\alpha \beta}_{{\bf Q}_\nu} {\tilde m}_{\alpha}({{\bf Q}_\nu}) {\tilde m}_{\beta}({{\bf Q}_\nu}))^2$) was required to stabilize the circular skyrmion phase (Phase IV), while it could not predict the appearance of Phases II or III. On the other hand, our present theoretical model (Eq. \ref{Hamiltonian}) suggests that the competition between RKKY interactions at inequivalent wave vectors ${\bf Q}_{\rm A}$ and ${\bf Q}_{\rm B}$ (Fig. 5a) is crucial for the appearance of Phases II, III and IV, and the four-spin interaction is not important to reproduce the magnetism in GdRu$_2$Ge$_2$. In this context, the observed meron/anti-meron and two distinctive skyrmion crystal phases are realized under the delicate balance between magnetic interactions mediated by itinerant electrons. (In Supplementary Note IX, the additional theoretical calculation has been performed to clarify the key difference between GdRu$_2$Si$_2$ and GdRu$_2$Ge$_2$.)

In this study, we reported the experimental discovery of multi-step topological transitions among the elliptic skyrmion, meron/anti-meron pair, and circular skyrmion phases in a centrosymmetric magnet GdRu$_2$Ge$_2$. Previously, the controlled transformation between different types of magnetic quasi-particles has been reported only for a few DM-based non-centrosymmetric magnets\cite{Meron1, Heusler_Peng, Heusler_Parkin_tilted, Bobber}, and it is remarkable that such a simple centrosymmetric compound can host even more intricate manner of topological magnetic transitions. The diameter of observed magnetic quasi-particles in GdRu$_2$Ge$_2$ is as small as 2.7 nm, which is one or two orders of magnitude smaller than traditional DM-based non-centrosymmetric compounds. Our theoretical analysis reveals that the competition between RKKY interactions at inequivalent wave vectors ${\bf Q}_{\rm A} = (q, 0, 0)$ and ${\bf Q}_{\rm B} = (q/2, q/2, 0)$, typically induced by the nesting of Fermi-surfaces along multiple directions\cite{HayamiModel1, Hayami_JPSJ}, is the key to realize a rich variety of nanometric particle-like spin textures in centrosymmetric systems. In principle, these magnetic quasi-particles will be transformable into each other by various external stimuli, because of their pseudo degeneracy and sizable energy barrier\cite{Oike}. Such a potential metastability may lead to the development of unique manner of multi-valued memory/logic function\cite{OlegReview, Bobber, SkBag}. Recent theoretical studies predict the appearance of even wider variety of nontrivial topological spin textures (such as higher-order skyrmion with $|N_{\rm sk}| \geq 2$) in centrosymmetric rare-earth compounds\cite{HayamiModel1,BatistaModel_Frustration,Hayami_JPSJ}. Further search for novel materials hosting exotic magnetic quasi-particles, as well as their direct real-space observation and manipulation, are the issue for the future study. (See Supplementary Note XI.)

\begin{figure}
\begin{center}
\includegraphics*[width=15.5cm]{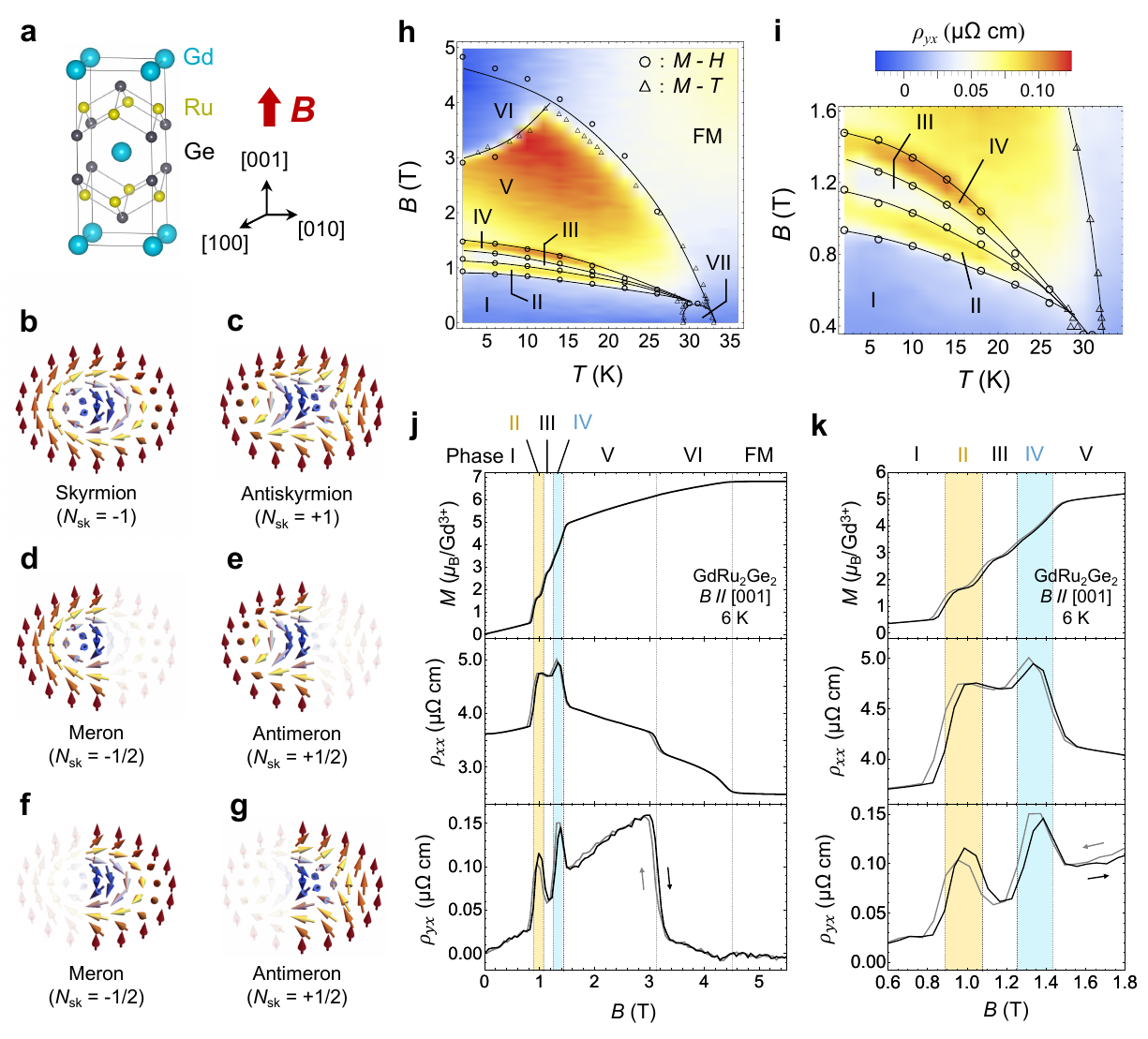}
\caption{{\bf Magnetic and electrical transport properties of GdRu$_2$Ge$_2$.} {\bf a}, Crystal structure of GdRu$_2$Ge$_2$. {\bf b}-{\bf g}, Schematic illustration of skyrmion, anti-skyrmion, meron and anti-meron spin textures. {\bf h},{\bf i}, $B$-$T$ magnetic phase diagram of GdRu$_2$Ge$_2$ for $B \parallel [001]$. The background color represents the value of Hall resistivity $\rho_{yx}$. Open circles (triangles) indicate phase boundaries obtained from the magnetic-field (temperature) dependence of magnetization $M$. FM represents the ferromagnetic state. {\bf j},{\bf k}, Magnetic-field dependence of magnetization $M$, longitudinal resistivity $\rho_{xx}$ and Hall resistivity $\rho_{yx}$ measured at 6 K for $B \parallel [001]$ and $I \parallel [100]$. Black and gray lines represent the field increasing and decreasing runs, respectively.
\label{Fig1}}
\end{center}
\end{figure}

\begin{figure}
\begin{center}
\includegraphics*[width=14.0cm]{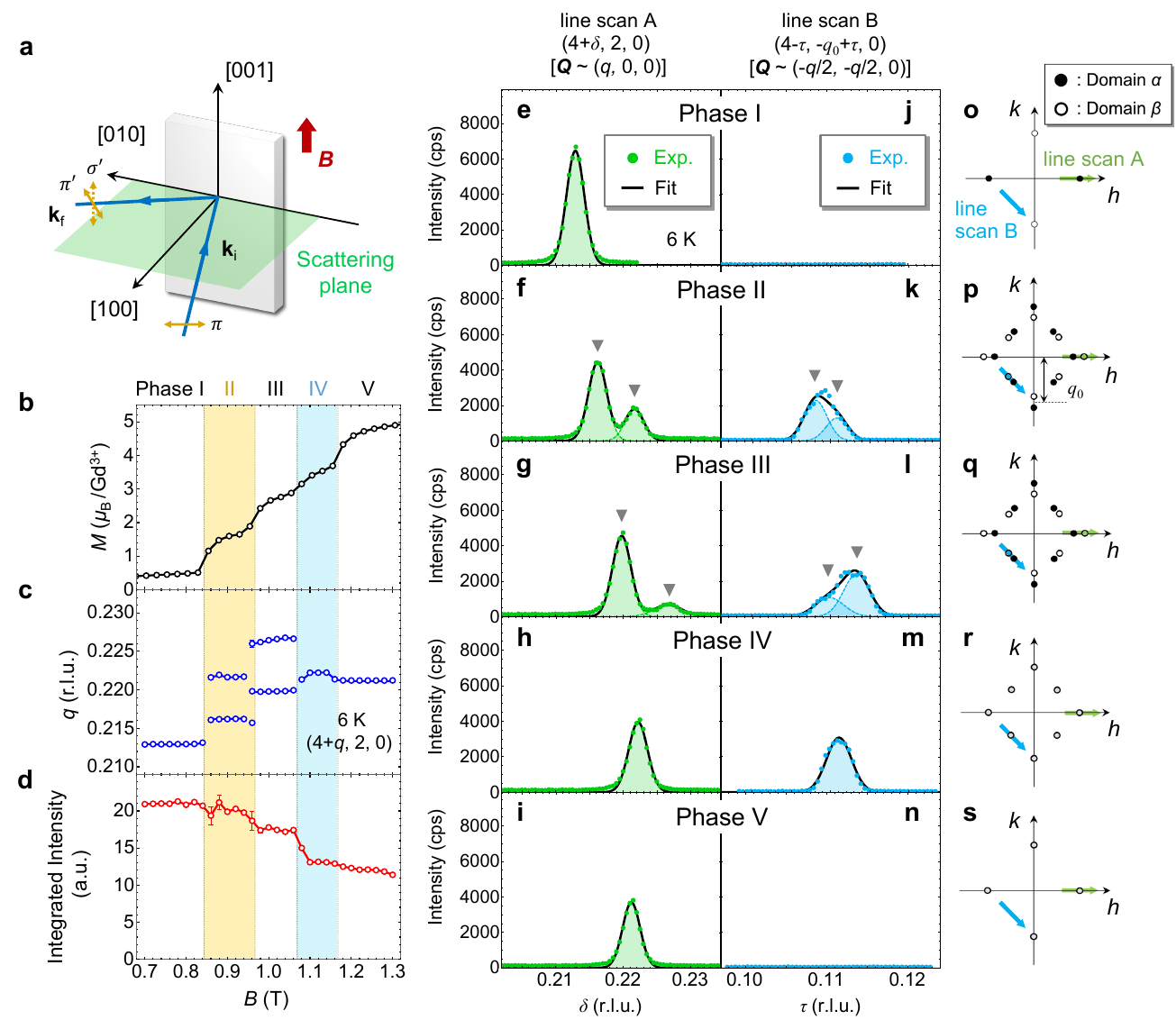}
\caption{{\bf RXS measurements for GdRu$_2$Ge$_2$.} ${\bf a}$, Schematic illustration of experimental setup for RXS measurement. The scattering plane spanned by the incident and the scattered beams (${\bf k}_i$ and ${\bf k}_f$, respectively) lies normal to the [001] axis. The incident X-ray beam is linearly polarized within the scattering plane ($\pi$-polarized). {\bf b}-{\bf d}, Magnetic-field dependence of magnetization $M$, the wave number $q$ and integrated intensity of the $(4+q, 2, 0)$ magnetic reflection for $B \parallel [001]$ at 6 K. {\bf e}-{\bf i}, Line profiles of $(4+\delta, 2, 0)$ scan (i.e. line scan A) for Phases I ($B = 0$), II (0.92 T), III (1.03 T), IV (1.13 T) and V (1.3 T) to identify the ${\bf Q} \sim (q, 0, 0)$ magnetic satellite peaks around the fundamental Bragg spot $(4, 2, 0)$. Each experimental data (closed circles) is fitted by one or two Gaussian functions. {\bf j}-{\bf n}, Line profiles of $(4-\tau,\ -q_0+\tau,\ 0)$ scan (i.e. line scan B) to identify ${\bf Q} \sim (q/2, q/2, 0)$ magnetic satellite peaks around the fundamental Bragg spot $(4, 0, 0)$ for Phases I-V. The definition of $q_0$ is shown in ({\bf p}). {\bf o}-{\bf s}, Reciprocal-space distribution of magnetic satellite reflections, as well as the directions of line scans A and B for Phases I-V. The selection rule of magnetic satellite reflections are common for both $(4, 2, 0)\pm{\bf Q}$ and $(4, 0, 0)\pm{\bf Q}$ positions. Closed and open circles in ${\bf o}$-${\bf q}$ represent the contributions of magnetic domains $\alpha$ and $\beta$, respectively.}
\label{Fig2}
\end{center}
\end{figure}

\begin{figure}
\begin{center}
\includegraphics*[width=15cm]{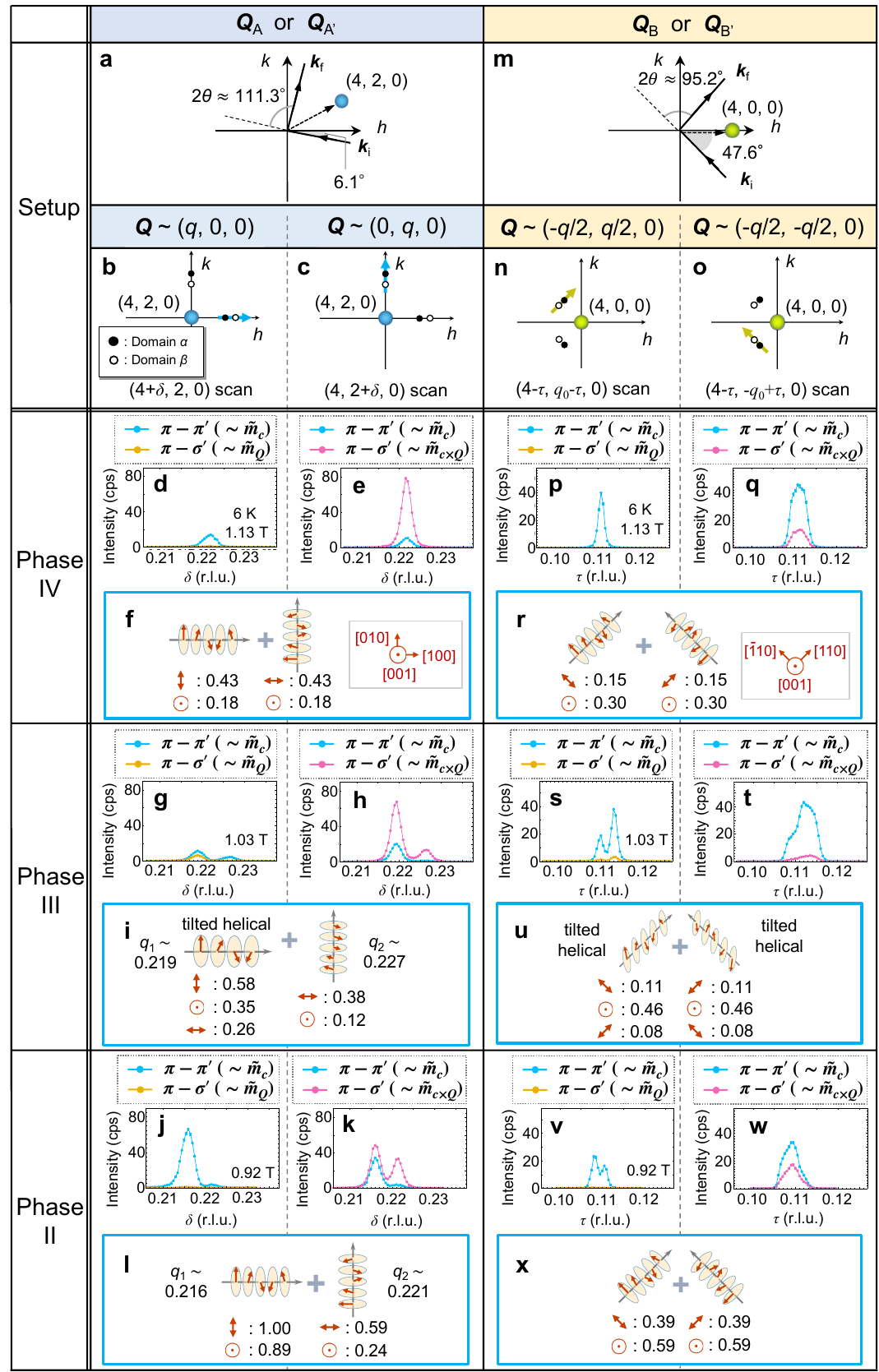}
\caption{{\bf Polarization analysis of RXS profiles in Phases II, III and IV.} All measurements were performed with the experimental setup shown in Fig. 2a, where the intensity of scattered X-ray with the polarization parallel and normal to scattered plane ($I_{\pi-\pi'}$ and $I_{\pi-\sigma'}$, respectively) was separately measured. According to Eq. (3), $I_{\pi-\pi'}$ always represents the $\tilde{m}_{c}$ component, and $I_{\pi-\sigma'}$ reflects the component of $\tilde{{\bf m}}({\bf Q})$ parallel to ${\bf k}_i$. 
{\bf a}, The measurement geometry for the magnetic satellites around the fundamental Bragg spot $(4, 2, 0)$. The incident vector ${\bf k}_{\rm i}$ is almost parallel to the [100] axis and thus $I_{\pi-\sigma'}$ should mainly reflect the [100] component of $\tilde{{\bf m}}({\bf Q})$.
{\bf b},{\bf c}, Schematic illustration of $(4+\delta, 2, 0)$ and $(4,\ 2+\delta,\ 0)$ line scans in the reciprocal space, to investigate the ${\bf Q} \sim (q, 0, 0)$ and $(0, q, 0)$ magnetic satellites around the fundamental Bragg spot $(4, 2, 0)$, respectively.
{\bf d},{\bf e}, $(4+\delta, 2, 0)$ and $(4,\ 2+\delta,\ 0)$ line profiles measured in Phase IV (1.13 T). The inset indicates the spin component $\tilde{m}_\alpha ({\bf Q})$ represented by $I_{\pi-\pi'}$ and $I_{\pi-\sigma'}$ for each scan. {\bf f}, Real-space schematic illustration of the modulated spin components $\tilde{m}_\alpha ({\bf Q})$ for the wave vectors ${\bf Q} = (q, 0, 0)$ and ${\bf Q}=(0, q, 0)$. The numbers represent the relative amplitude of $\tilde{m}_\alpha ({\bf Q})$ along the direction denoted by red arrow (i.e. $\alpha = c$, $Q$ or $c \times Q$). 
{\bf g}-{\bf i} and {\bf j}-{\bf l}, The corresponding data measured at Phases III (1.03 T) and II (0.92 T), respectively. 
{\bf m}, The measurement geometry for the magnetic satellites around the fundamental Bragg spot $(4, 0, 0)$. The incident vector ${\bf k}_{\rm i}$ is almost parallel to the [1$\bar{1}$0] axis and thus $I_{\pi-\sigma'}$ should mainly reflect the [1$\bar{1}$0] component of $\tilde{{\bf m}}({\bf Q})$.
{\bf n},{\bf o}, Schematic illustration of $(4-\tau, q_0-\tau, 0)$ and $(4-\tau, -q_0+\tau, 0)$ line scans in the reciprocal space, to investigate the ${\bf Q} \sim (-q/2, q/2, 0)$ and $(-q/2, -q/2, 0)$ magnetic satellites around the fundamental Bragg spot $(4, 0, 0)$, respectively. {\bf p},{\bf q}, $(4-\tau, q_0-\tau, 0)$ and $(4-\tau, -q_0+\tau, 0)$ line profiles measured in Phase IV (1.13 T). {\bf r}, Real-space schematic illustration of the modulated spin components $\tilde{m}_\alpha ({\bf Q})$ for the wave vectors ${\bf Q} \sim (-q/2, q/2, 0)$ and ${\bf Q} \sim (-q/2, -q/2, 0)$. The numbers represent the relative amplitude of $\tilde{m}_\alpha ({\bf Q})$ along the direction denoted by red arrow. {\bf s}-{\bf u} and {\bf v}-{\bf x}, The corresponding data measured at Phases III (1.03 T) and II (0.92 T), respectively.}

\label{Fig3}
\end{center}
\end{figure}

\begin{figure}
   \begin{center}
   \includegraphics*[width=17cm]{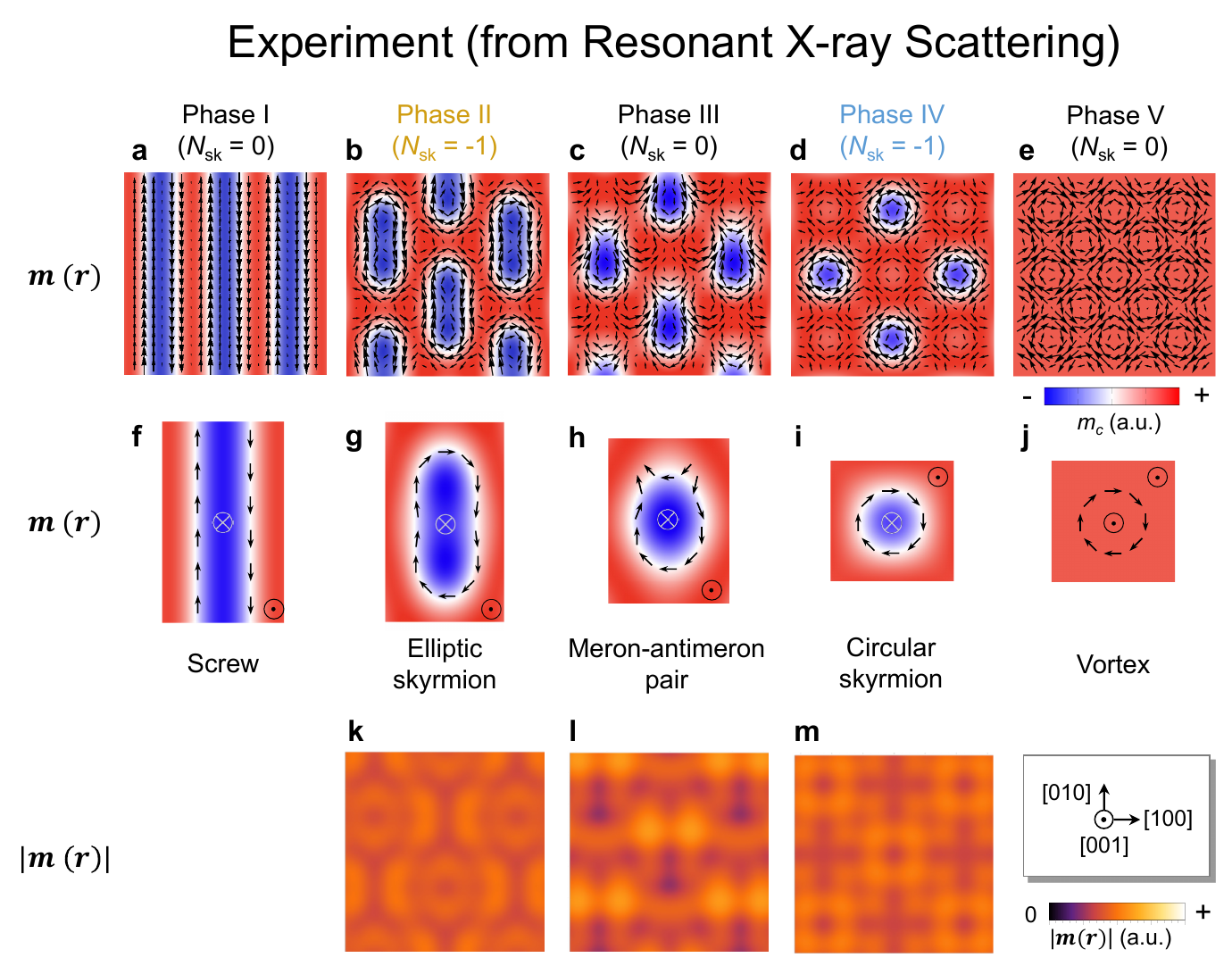}
   \caption{{\bf Experimentally deduced magnetic structures for GdRu$_2$Ge$_2$.} {\bf a}-{\bf e}, The magnetic structure ${\bf m} ({\bf r})$ for Phases I, II, III, IV and V, reconstructed based on Eq. (\ref{ReconstructEq}) and $\tilde{m}_\alpha ({\bf Q}_\nu)$ (${\bf Q}_\nu = {\bf Q}_{\rm A}$, ${\bf Q}_{\rm A'}$, ${\bf Q}_{\rm B}$, ${\bf Q}_{\rm B'}$, ${\bf Q}_{\rm C}$ and ${\bf Q}_{\rm C'}$) deduced by RXS experiments as summarized in Supplementary Table 1. The relative phase $\theta^{{\bf Q}_\nu}_\alpha$ is determined so as to realize the most uniform $|{\bf m} ({\bf r})|$ distribution (See Supplementary Note III for the detail). The black arrows and background color represent the in-plane and out-of-plane component of local magnetic moment ${\bf m}({\bf r})$, respectively. {\bf f}-{\bf j}, Schematic illustration of spin textures in {\bf a}-{\bf e}. {\bf k}-{\bf m}, The spatial distribution of $|{\bf m} ({\bf r})|$ for the spin textures in {\bf b}-{\bf d}. The small amount of non-uniform component remaining in the $|{\bf m}({\bf r})|$ profile is associated with the higher-order harmonics neglected in the present model.}
   \label{Fig4}
   \end{center}
\end{figure}

\begin{figure}
\begin{center}
\includegraphics*[width=16cm]{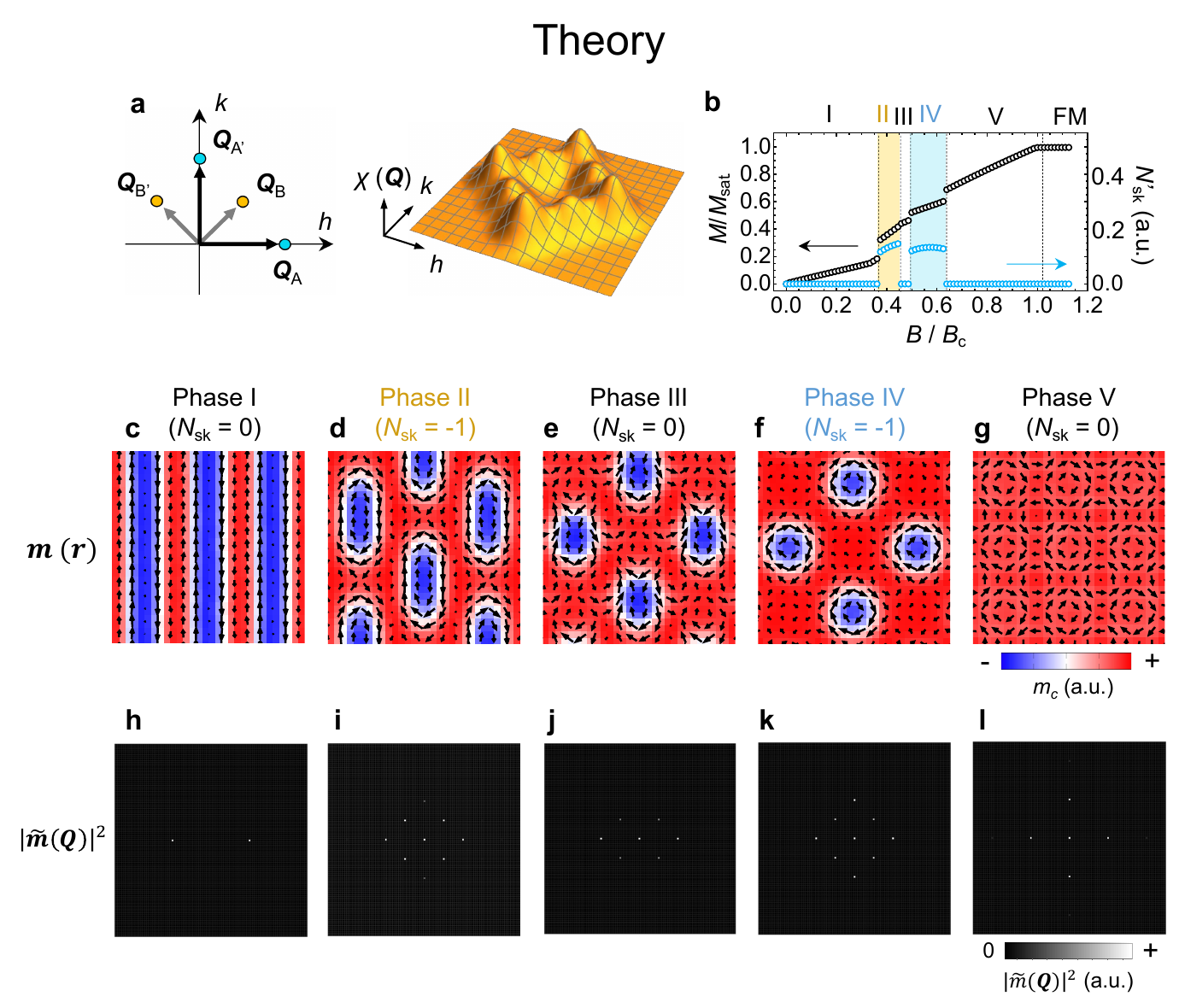}
\caption{{\bf Theoretical magnetic structures obtained by simulated annealing.} {\bf a}, Schematic illustration of bare-susceptibility $\chi ({\bf Q})$ distribution considered in the effective magnetic Hamiltonian Eq. (6), which assumes the largest peaks at ${\bf Q}_{\rm A} = (q, 0)$ and ${\bf Q}_{\rm A'} = (0, q)$ with $q=2\pi/5$ as well as the relatively large value at ${\bf Q}_{\rm B} = (q/2, q/2)$ and ${\bf Q}_{\rm B'} = (-q/2, q/2)$. {\bf b}, Magnetic-field dependence of magnetization $M$ and scalar spin chirality $N'_{\rm sk}$ defined in Methods section, obtained by the simulated annealing with magnetic field $B$ applied normal to the square lattice. $M_{\rm sat}$ and $B_{\rm c}$ represent the saturated magnetization and the critical magnetic field to obtain fully polarized ferromagnetic (FM) state, respectively. {\bf c}-{\bf g}, Theoretically simulated magnetic structure ${\bf m} ({\bf r})$ in Phases I, II, III, IV and V, obtained at $B = 0$, $0.39 B_c$, $0.47 B_c$, $0.56 B_c$ and $0.67 B_c$, respectively. {\bf h}-{\bf l}, The corresponding reciprocal-space distribution of $|\tilde{{\bf m}}({\bf Q})|^2$, which is expected to scale with the scattering intensity at each ${\bf Q}$ position in the RXS experiments. See Methods section and Supplementary Note VIII for the detail of theoretical calculations.}
\label{Fig5}
\end{center}
\end{figure}

\section*{Methods}

\subsection*{Sample preparation and characterization.}

Polycrystalline rods of GdRu$_{2}$Ge$_{2}$ were prepared by the arc-melting technique from stoichiometric amount of pure Gd, Ru, and Ge pieces using a water-cooled copper crucible under an Ar atmosphere. Bulk single crystals were grown in Ar gas flow by using a floating zone furnace. The crystal orientation was determined using the X-ray Laue method, and the phase purity of the samples was confirmed by the powder X-ray diffraction.

\subsection*{Magnetic and electrical transport property measurements.}

Magnetization measurements (Figs. 1j, 1k and 2b in the main text and the top panels of Supplementary Figs. 8b and 11c) on rectangular-shaped polished samples were performed using a Magnetic Properties Measurement System (MPMS, Quantum Design) with SQUID (superconducting quantum interference device) magnetometer. Measurements of the electrical transport properties were performed with the conventional five-terminal method using the AC-transport option in a Physical Properties Measurement System (PPMS, Quantum Design). The same sample was used for the measurements of magnetization and electrical transport properties in Fig. 1. Since the Hall resistivity (longitudinal resistivity) is an odd (even) function of magnetic field $B$, the measured $\rho_{yx}$-$B$ profile was anti-symmetrized with respect to $B$ to eliminate the possible contamination of the longitudinal resistivity component\cite{THE}. Note that the shape of the sample used for the resonant X-ray scattering experiment in Figs. 2 and 3 was different from the one used for the electrical transport measurements in Fig. 1, which caused the slight discrepancy of critical $B$ value via the shape anisotropy associated with the demagnetizing field. In the present work, the typical thickness of GdRu$_2$Ge$_2$ sample is in order of 1 mm. We investigated several crystal pieces with different thickness, and the multi-step topological transition was commonly observed for all samples.

\subsection*{Neutron scattering measurements.}

Neutron scattering measurement was performed using the high-resolution chopper (HRC) spectrometer at BL12 \cite{HRC} in the Materials and Life Science Facility (MLF) of Japan Proton Accelerator Research Complex (J-PARC). %
A single crystal with a polished flat (100) plane (dimensions of 6.5$\times$8.0$\times$0.4 mm) was loaded into a vertical-field superconducting magnet. %
The field direction was parallel to the [001] axis of the sample. %
To reduce the strong neutron absorption effect of Gd, the energy of the incident neutron beam was tuned to 153.5 meV by using a high-resolution Fermi chopper. %
The energy resolution at the elastic condition was approximately 4\%. 
The intensities of the scattered neutrons were measured by arrays of $^3$He-position sensitive detectors. %
The measurements were repeated with different $\omega$ angles (rotation about the [001] axis). The data were processed by DAVE software\cite{DAVE} to obtain the intensity distributions in the reciprocal lattice space shown in Supplementary Fig. 8. 

\subsection*{Resonant X-ray scattering measurements.}

RXS measurement was performed at BL-3A, Photon Factory, KEK, Japan. 
The photon energy was adjusted in resonance with the Gd $L_{2}$ absorption edge ($\sim$7.935 keV), as detailed in Supplementary Fig. 7 and Supplementary Note VI.
A single crystal with a polished flat (100) plane (dimensions of 0.45$\times$3.2$\times$4.6 mm) was attached on an Al plate with varnish and loaded into a vertical-field superconducting magnet, so that the magnetic field was applied parallel to the [001] axis. (The accuracy of magnetic-field value is within 0.01 T.) 
The incident X-ray beam was horizontally polarized and the scattering plane was perpendicular to the [001] axis. To analyze the polarization of the scattered X-ray beam, we used the 006 reflection of a pyrolytic graphite plate, where the 2$\theta$ angle for the analyzer at the Gd $L_2$ edge was 88.7 degrees.  By rotating the pyrolytic graphite plate about the scattered beam, the $\sigma$' ($\pi$') component, polarized perpendicular (parallel) to the scattering plane, was selectively detected. We also performed measurements without analyzing polarization of the scattered X-ray beam, where the scattered beam included both the $\sigma$'- and $\pi$'-polarized  components. In Figs. 2c and d, the error bars correspond to the asymptotic standard errors in the least-squares fitting analysis of experimental data in Figs. 2e-i using the Gaussian functions. The possibility of double scatterings for the observed higher-order peaks could be ruled out because of the absence of the peaks in the multi-domain state of Phase I.

\subsection*{Theoretical calculation.}

The data in Fig.~\ref{Fig5} were obtained by performing numerical calculations based on simulated annealing for the model in Eq.~(\ref{Hamiltonian}) on a square lattice. 
The temperature was gradually reduced from $T=1$ to $0.01$ with a condition $T_{n+1}=\alpha T_{n}$, where $T_n$ was the temperature in the $n$th step and $\alpha$ was set between $0.99999$ and $0.999999$. 
$10^5$-$10^6$ Monte Carlo steps were performed for thermalization and measurements followed by the standard Metropolis algorithm. 
In the vicinity of the phase boundaries, the simulations were also performed from the spin patterns obtained at low temperatures. 
The model parameters in Eq.~(\ref{Hamiltonian}) were set as $J=1$, $\kappa=0.86$, $\gamma_1=0.9$, $\gamma_2=0.91675$, and $\gamma_3=0.0725$ for 
$\Gamma^{xx}_{\mathbf{Q}_{\rm A}}=\Gamma^{yy}_{\mathbf{Q}_{\rm A'}}=\gamma_1\gamma_2$, 
$\Gamma^{yy}_{\mathbf{Q}_{\rm A}}=\Gamma^{xx}_{\mathbf{Q}_{\rm A'}}=\gamma_1$, 
$\Gamma^{zz}_{\mathbf{Q}_{\rm A}}=\Gamma^{zz}_{\mathbf{Q}_{\rm A'}}=1$, 
$\Gamma^{xx}_{\mathbf{Q}_{\rm B}}=\Gamma^{yy}_{\mathbf{Q}_{\rm B}}=\Gamma^{xx}_{\mathbf{Q}_{\rm B'}}=\Gamma^{yy}_{\mathbf{Q}_{\rm B'}}=\kappa \gamma_1$,
$-\Gamma^{xy}_{\mathbf{Q}_{\rm B}}=-\Gamma^{yx}_{\mathbf{Q}_{\rm B}}=\Gamma^{xy}_{\mathbf{Q}_{\rm B'}}=\Gamma^{yx}_{\mathbf{Q}_{\rm B'}}=\kappa \gamma_3$,
for $\Gamma^{zz}_{\mathbf{Q}_{\rm B}}=\Gamma^{zz}_{\mathbf{Q}_{\rm B'}}=\kappa$ to satisfy four-fold rotational symmetry of the square lattice (the other components of $\Gamma^{\alpha\beta}_{\mathbf{Q}_\nu}$ were zero), where $x=[100]$, $y=[010]$, and $z=[001]$; 
$\gamma_1<1$ represents the easy-axis anisotropy, while $\gamma_2$ and $\gamma_3$ represent the in-plane bond-dependent anisotropy that fixes the spiral plane; $\kappa<1$ represents the competition between interactions at different wave vectors to hold the relation $\chi({\bf Q}_{\rm A}) = \chi({\bf Q}_{\rm A'}) > \chi({\bf Q}_{\rm B}) = \chi({\bf Q}_{\rm B'})$. 
The system size was set as $N=100^2$. 
The scalar spin chirality $N'_{\rm sk}$ in Fig.~\ref{Fig5}b was calculated as 
\begin{equation}
N'_{\rm sk}=\bigg[ \frac{1}{N}
\sum_{i,\delta=\pm1}\mathbf{m}(\mathbf{r}_i) \cdot \left\{\mathbf{m}(\mathbf{r}_i+\delta\hat{\mathbf{e}}_x) \times \mathbf{m}(\mathbf{r}_i+\delta\hat{\mathbf{e}}_y) \right\}\bigg]^2, 
\end{equation}
where $\hat{\mathbf{e}}_x$ ($\hat{\mathbf{e}}_y$) is the unit vector in the $x$ ($y$) direction on the square lattice. The scalar spin chirality $N'_{\rm sk}$ represents the non-coplanarity of spin texture and usually scales with the amplitude of topological Hall signal $\rho_{yx}^T$\cite{RMP_AHE}, while the value is not quantized unlike $N_{\rm sk}$ defined in the continuum model, Eq. (1).

\section*{Data availability}

The data presented in the current study are available from the corresponding authors on reasonable request.

\section*{Author contributions}

S.S. and R.T. conceived the project. H.Y. and R.T. grew single crystals and characterized the magnetic and transport properties with the assistance of S.S., N.D.K. and Y.T. RXS measurements were carried out by H.Y., R.T., J.J. and T.N. with the assistance of H.Sai., H.Sag., H.N. and T.A. Neutron scattering measurements were carried out by J.J., H.Sai., S.I. and T.N. Theoretical simulation was performed by S.H. The manuscript was written by H.Y. and S.S. with the assistance of S.H. and T.N. All the authors discussed the results and commented on the manuscript.

\section*{Acknowledgements} The authors thank Y. Motome, A. Kikkawa, X. Z. Yu, N. Shibata, T. Seki, and S. Toyama for enlightening discussions and experimental helps. This work was partly supported by Grants-In-Aid for Scientific Research (grant nos. 18H03685, 20H00349, 21H04440, 21H04990, 21K13876, 21K18595, 22H04965, 22H04468, 22KJ1061, 23K13069, 23H04869) from JSPS, PRESTO (grant nos JPMJPR18L5, JPMJPR20B4, JPMJPR20L8) and CREST (grant no. JPMJCR1874, JPMJCR23O4) from JST, Katsu Research Encouragement Award and UTEC-UTokyo FSI Research Grant Program of the University of Tokyo, Asahi Glass Foundation and Murata Science Foundation. This work is based on experiments performed at Materials and Life Science Experimental Facility (MLF) in Japan Proton Accelerator Research Complex (J-PARC) (Proposal No. 2020S01), and Photon Factory in High Energy Accelerator Research Organization (Proposal No. 2020G665). The illustration of crystal structure was drawn by VESTA\cite{VESTA}.

\section*{Additional information}
Supplementary Information is available in the online version of the paper. Correspondence and requests for materials should be addressed to H.Y. and S.S.
\section*{Competing financial interests}
The authors declare that they have no competing financial interests.

\newpage

\end{document}


\title{Supplementary Materials: \\Multi-step topological transitions among meron and skyrmion crystals in a centrosymmetric magnet}

\author{H. Yoshimochi$^{1}$, R. Takagi$^{1,2,3}$, J. Ju$^{4}$, N. D. Khanh$^{5}$, H. Saito$^{4}$, H. Sagayama$^{6}$, H. Nakao$^{6}$, S. Itoh$^{6,7}$, Y. Tokura$^{1, 5, 8}$, T. Arima$^{5, 9}$, S. Hayami$^{10}$, T. Nakajima$^{4, 5}$, S. Seki$^{1, 2, 3}$}

\affiliation{$^1$ Department of Applied Physics, University of Tokyo, Tokyo 113-8656, Japan, \\ 
$^2$ Institute of Engineering Innovation, University of Tokyo, Tokyo 113-0032, Japan, \\ 
$^3$ PRESTO, Japan Science and Technology Agency (JST), Kawaguchi 332-0012, Japan, \\ 
$^4$ Institute for Solid State Physics, University of Tokyo, Kashiwa 277-0882, Japan,\\
$^5$ RIKEN Center for Emergent Matter Science (CEMS), Wako 351-0198, Japan,\\
$^6$ Institute of Materials Structure Science, High Energy Accelerator Research Organization, Tsukuba 319-1195, Japan,\\
$^7$ Materials and Life Science Division, J-PARC Center, Tokai, Ibaraki 319-1195, Japan, \\
$^8$ Tokyo College, University of Tokyo, Tokyo 113-8656, Japan,\\
$^9$ Department of Advanced Materials Science, University of Tokyo, Kashiwa 277-8561, Japan,\\
$^{10}$ Graduate School of Science, Hokkaido University, Sapporo 060-0810, Japan}

\maketitle
\color{black}

\section{Polarization analysis of resonant X-ray scattering profiles in Phases I and V}

As discussed in the main text, the spin textures in Phases I and V contain the spin component modulated with ${\bf Q}_{\rm A} = (q, 0, 0)$, but not the one with  ${\bf Q}_{\rm B} = (q/2, q/2, 0)$ (Fig. 2 in the main text). To investigate the detailed spin orientations for these phases, the polarization of scattered X-ray has been analyzed with the same experimental setup as the one employed for Phases II - IV (Fig. 2a in the main text). 

\begin{figure}[p]
\begin{center}
\includegraphics*[width=12.4cm]{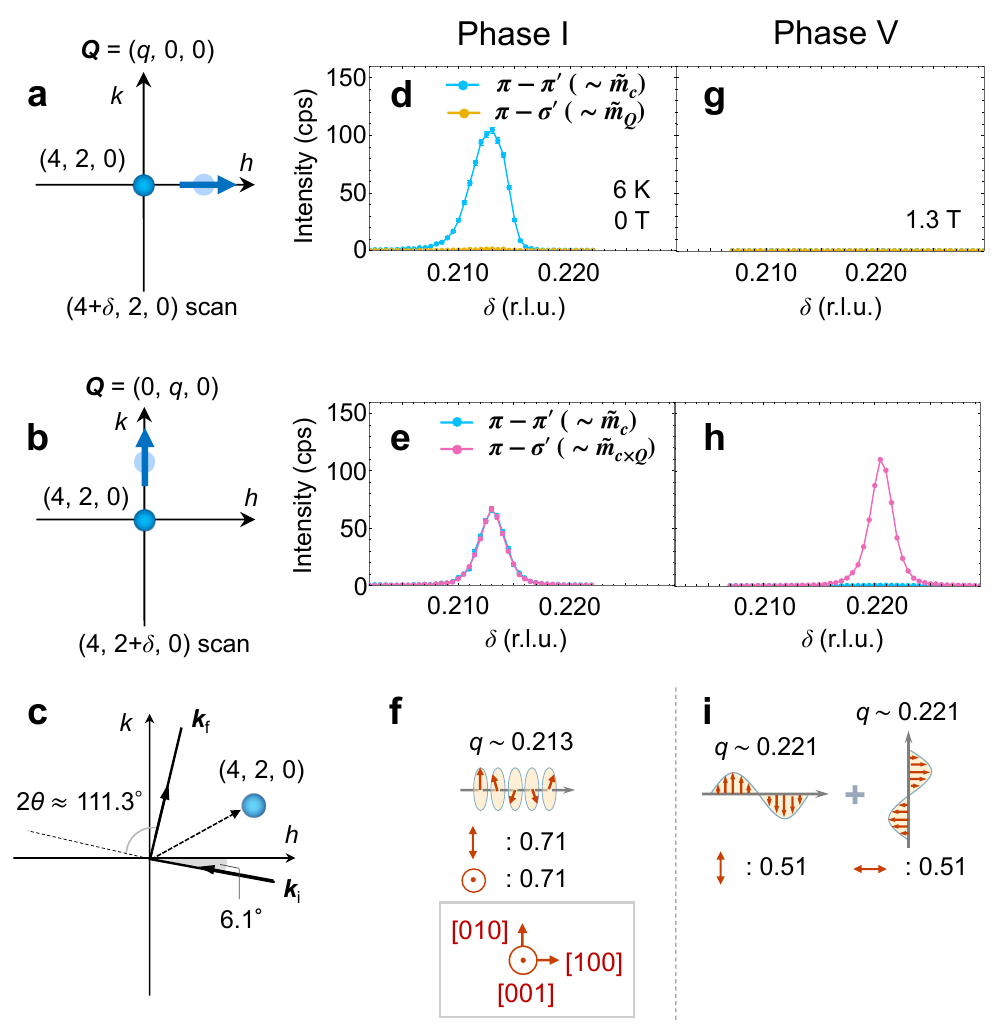}
\caption{{\bf Polarization analysis of RXS profiles in Phases I and V.} All measurements were performed with the experimental setup shown in Fig. 2a in the main text, where the intensity of scattered X-ray with the polarization parallel and normal to scattered plane ($I_{\pi-\pi'}$ and $I_{\pi-\sigma'}$, respectively) was separately measured. According to Eq. (3) in the main text, $I_{\pi-\pi'}$ always represents the $\tilde{m}_{c}$ component, and $I_{\pi-\sigma'}$ reflects the component of $\tilde{{\bf m}}({\bf Q})$ parallel to ${\bf k}_i$. {\bf a}, {\bf b}, Schematic illustration of $(4+\delta, 2, 0)$ and $(4,\ 2+\delta,\ 0)$ line scan in the reciprocal space, to investigate the ${\bf Q} \sim (q, 0, 0)$ and $(0, q, 0)$ magnetic satellites around the fundamental Bragg spot $(4, 2, 0)$, respectively. The associated measurement geometry is illustrated in ({\bf c}), where the incident vector ${\bf k}_{\rm i}$ is nearly parallel to the [100] axis and thus $I_{\pi-\sigma'}$ should mainly reflect the [100] component of $\tilde{{\bf m}}({\bf Q})$. {\bf d},{\bf e}, The $(4+\delta, 2, 0)$ and $(4,\ 2+\delta,\ 0)$ line profiles measured in Phase I ($B=0$). The inset indicates the spin component $\tilde{m}_\alpha ({\bf Q})$ represented by $I_{\pi-\pi'}$ and $I_{\pi-\sigma'}$ for each scan. {\bf f}, Real-space schematic illustration of the modulated spin components $\tilde{m}_\alpha ({\bf Q})$ for the wave vector ${\bf Q} = (q, 0, 0)$. The numbers represent the relative amplitudes of $\tilde{m}_\alpha ({\bf Q})$ along the direction denoted by red arrows (i.e. $\alpha = c$, $Q$, or $c \times Q$). {\bf g}-{\bf i}, The corresponding data measured in Phase V (1.3 T).}
\end{center}
\end{figure}

For this purpose, we investigated the magnetic satellites $(4, 2, 0) \pm {\bf Q}$ with the measurement geometry shown in Supplementary Fig. 1c (i.e. the same as Fig. 3a in the main text). In this configuration, ${\bf k}_i$ is nearly parallel to the [100] axes, in which $I_{\pi-\sigma'}$ and $I_{\pi-\pi'}$ mainly reflect the [100] and [001] components of $\tilde{{\bf m}}({\bf Q})$, respectively. Supplementary Figs. 1a and d show the ($4+\delta$, 2, 0) line profile in Phase I, where the presence of $I_{\pi-\pi'}$ (absence of $I_{\pi-\sigma'}$) peak suggests the presence of $\tilde{m}_{c}$ (absence of $\tilde{m}_{Q}$) for ${\bf Q}_{\rm A} = (q, 0, 0)$. The corresponding (4, $2+\delta$, 0) line profile is shown in Supplementary Figs. 1b and e, where the presence of $I_{\pi-\pi'}$ and $I_{\pi-\sigma'}$ indicate the presence of $\tilde{m}_{c}$ and $\tilde{m}_{c\times Q}$ for ${\bf Q}_{\rm A'} = (0, q, 0)$. These results suggest that Phase I represents the screw spin state, as shown in Supplementary Fig. 1f, where their neighboring spins rotate within a plane normal to the wave vector.

Similar measurements were also performed for Phase V. The results are summarized in Supplementary Figs. 1g and h. In this case, the absence of $\tilde{m}_{c}$ and $\tilde{m}_{Q}$ for ${\bf Q}_{\rm A} = (q, 0, 0)$, as well as the presence of $\tilde{m}_{c \times Q}$ and the absence of $\tilde{m}_{c}$ for ${\bf Q}_{\rm A'} = (0, q, 0)$, are confirmed. It suggests that ${\bf Q}_{\rm A}$ and ${\bf Q}_{\rm A'}$ in Phase V are characterized by sinusoidal modulation of in-plane spin component normal to the wave vector, as shown in Supplementary Fig. 1i. If we assume Phase V as a double-${\bf Q}$ state, their superposition deduced from Eq. (5) in the main text represents the square vortex lattice state, as shown in Figs. 4e in the main text. 

In the theoretical simulation, such a square vortex lattice state is stabilized in a high out-of-plane magnetic field (Fig. 5g in the main text). The corresponding $|\tilde{{\bf m}} ({\bf Q})|^2$ profile (Fig. 5l in the main text) predicts the presence and absence of RXS peak at ${\bf Q}_{\rm A} = (q, 0, 0)$ and ${\bf Q}_{\rm B} = (q/2, q/2, 0)$ position, respectively, which is consistent with the experimental results obtained for Phase V (Figs. 2i, n and s in the main text). In case of the isostructural GdRu$_2$Si$_2$, such a square vortex lattice state has indeed been identified by the real-space STM (scanning tunneling microscopy) measurements\cite{GdRu2Si2_STM} in a high magnetic field $B \parallel [001]$. On the basis of these analyses, we conclude that the Phase V in the present GdRu$_2$Ge$_2$ represents the double-${\bf Q}$ square vortex lattice state.

Note that the square vortex lattice state (Figs. 4e in the main text) is characterized by the vortex and anti-vortex arrangement of in-plane spin component, as well as the spatially uniform out-of-plane spin component. In this case, the local spin texture can wrap only less than a half of sphere (i.e. $|N_{\rm sk}| < 1/2$), and this phase cannot be considered as the meron/anti-meron lattice.  

\section{Magnetic domains}

\begin{figure}[b]
\begin{center}
\includegraphics*[width=15cm]{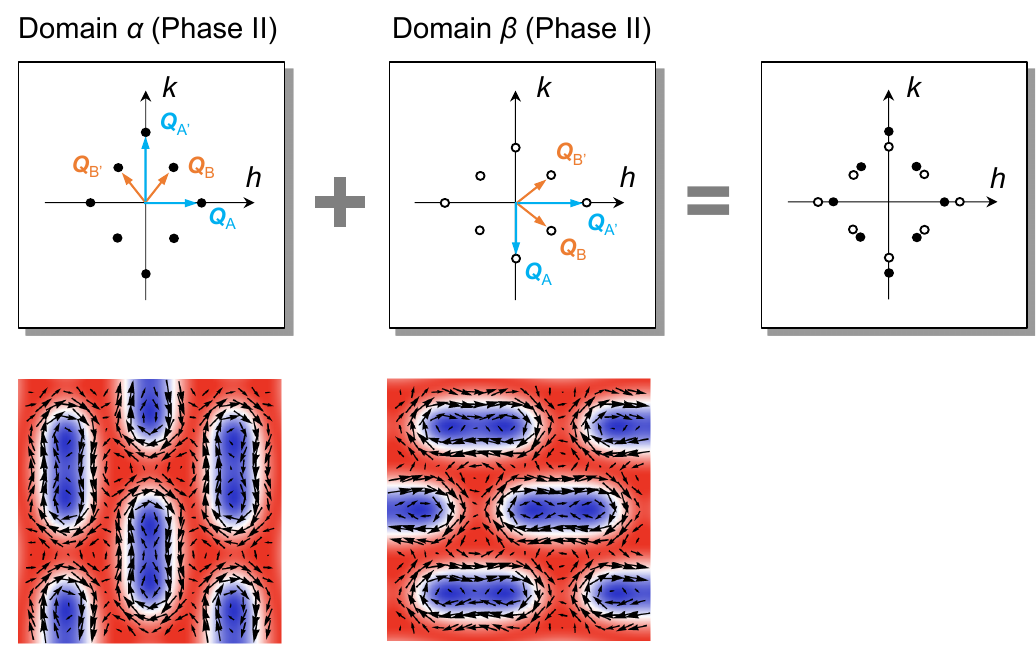}
\caption{{\bf Magnetic domains in Phase II.} Left (middle) panel indicates the reciprocal-space distribution of magnetic satellites and the corresponding real-space spin texture ${\bf m}({\bf r})$ for domain $\alpha$ ($\beta$). Right panel indicates the overall reciprocal-space magnetic scattering patten expected from the coexistence of domains $\alpha$ and $\beta$, which corresponds to Fig. 2p in the main text.}
\end{center}
\end{figure}

As discussed in the main text, the four-fold symmetry is broken in Phases II and III. When the magnetic structure breaks the symmetry of the original crystal structure, there should appear multiple equivalent magnetic domains that are converted into each other by the broken symmetry elements. In the present case, there should appear at least two magnetic domains $\alpha$ and $\beta$ related by four-fold rotation, as shown in Supplementary Fig. 2. Domain $\alpha$ is characterized by the ${\bf Q}_{\rm A} = (q_1, 0, 0)$, ${\bf Q}_{\rm A'} = (0, q_2, 0)$, ${\bf Q}_{\rm B} = (q_1/2, q_2/2, 0)$ and ${\bf Q}_{\rm B'} = (-q_1/2, q_2/2, 0)$, while the domain $\beta$ is characterized by the ${\bf Q}_{\rm A} = (0, -q_1, 0)$, ${\bf Q}_{\rm A'} = (q_2, 0, 0)$, ${\bf Q}_{\rm B} = (q_2/2, -q_1/2, 0)$ and ${\bf Q}_{\rm B'} = (q_2/2, q_1/2, 0)$. In both cases, the relations ${\bf Q}_{\rm A} = {\bf Q}_{\rm B} - {\bf Q}_{\rm B'}$ and ${\bf Q}_{\rm A'} = {\bf Q}_{\rm B} + {\bf Q}_{\rm B'}$ are satisfied. These two magnetic domains are degenerate in energy for $B \parallel [001]$ and expected to coexist in nearly equal populations. It leads to the reciprocal-space magnetic scattering pattern, as shown in the right panel of Supplementary Fig. 2, which corresponds to Figs. 2p and q in the main text.

Note that the crystal structure of the present compound is centrosymmetric, while the spin textures in Phases I - IV break the space inversion symmetry. It leads to the appearance of at least two possible helicity domains that are converted into each other by the space-inversion operation. For example, in the skyrmion lattice phases (Phases II and IV), the right-handed and left-handed skyrmion domains, characterized by the clockwise and counter-clockwise swirling manner of in-plane spin component, will coexist with the equal population ratio. These helicity domains produce the same magnetic scattering pattern, and therefore do not affect the analysis of RXS results.

\section{Reconstruction of real-space spin texture based on the RXS data}

\begin{table}[b]

        \caption{Modulated spin components ${\tilde m}_{\alpha} ({{\bf Q}_\nu})$ in Phases II, III, IV and V, deduced from the RXS experiments performed at 6 K. Here, we assume $q_1 < q_2$ for Phases II and III and $q_1 = q_2$ for Phases IV and V. $M_c^{\rm sat}$ is the saturated $M$ for $B \parallel [001]$.}
        
    \centering
     
     \begin{tabular}{|c|c|c|c|c|} \hline 
     
     ${\bf Q}$ &
     \begin{tabular}{c}
         ${\bf Q}_{\rm A} = (q_1,0,0)$ \\ ${\bf Q}_{\rm A'} = (0,q_2,0)$
     \end{tabular}
     &    
     \begin{tabular}{c}
         ${\bf Q}_{\rm B} = (q_1/2,q_2/2,0)$ \\ ${\bf Q}_{\rm B'} = (-q_1/2,q_2/2,0)$
     \end{tabular}
     &
     \begin{tabular}{c}
         ${\bf Q}_{\rm C} = (q_1,q_2,0)$ \\ ${\bf Q}_{\rm C'} = (-q_1,q_2,0)$
     \end{tabular}
     &
     \begin{tabular}{c}
         $(0,0,0)$
     \end{tabular}\\ \hline \hline
          
     Phase II& 
     \begin{tabular}{c}
         $\tilde{m}_{c \times Q}({{\bf Q}_{\rm A}}) =1.00$ \\ $\tilde{m}_{c}({{\bf Q}_{\rm A}}) =0.89$ \\ $\tilde{m}_{c \times Q}({{\bf Q}_{\rm A'}}) = 0.59$ \\ $\tilde{m}_{c}({{\bf Q}_{\rm A'}}) =0.24$
     \end{tabular}
     &    
     \begin{tabular}{c}
         $\tilde{m}_{c \times Q} =0.39$ \\ $\tilde{m}_{c} =0.59$ 
     \end{tabular}
     &
     \begin{tabular}{c}
         $\tilde{m}_{c \times Q} =0.08$ \\ $\tilde{m}_{c} =0.15$ 
     \end{tabular}
     &
     \begin{tabular}{c}
         $\tilde{m}_{c}^0 =0.23M_{c}^{\rm sat.}$
     \end{tabular}\\ \hline

     Phase III& 
     \begin{tabular}{c}
          $\tilde{m}_{c \times Q}({{\bf Q}_{\rm A}}) =0.58$ \\$\tilde{m}_{c}({{\bf Q}_{\rm A}}) =0.35$ \\ $\tilde{m}_{Q}({{\bf Q}_{\rm A}}) =0.26$ \\ $\tilde{m}_{c \times Q}({{\bf Q}_{\rm A'}}) = 0.38$ \\ $\tilde{m}_{c}({{\bf Q}_{\rm A'}}) =0.12$ 
     \end{tabular}
     &    
     \begin{tabular}{c}
         $\tilde{m}_{c \times Q} =0.11$ \\ $\tilde{m}_{c} =0.46$ \\$\tilde{m}_{Q} =0.08$ 
     \end{tabular}
     &
     \begin{tabular}{c}
        ---
     \end{tabular}
     &
     \begin{tabular}{c}
         $\tilde{m}_{c}^0 =0.41M_{c}^{\rm sat.}$
     \end{tabular}\\ \hline
     
     Phase IV& 
     \begin{tabular}{c}
         $\tilde{m}_{c \times Q} = 0.43$ \\ $\tilde{m}_{c} =0.18$ 
     \end{tabular}
     &    
     \begin{tabular}{c}
         $\tilde{m}_{c \times Q} =0.15$ \\ $\tilde{m}_{c} =0.30$
     \end{tabular}
     &
     \begin{tabular}{c}
         $\tilde{m}_{c \times Q} =0.03$ \\ $\tilde{m}_{c} =0.07$
     \end{tabular}
     &
     \begin{tabular}{c}
         $\tilde{m}_{c}^0 =0.57M_{c}^{\rm sat.}$
     \end{tabular}\\ \hline
     
     Phase V& 
     \begin{tabular}{c}
         $\tilde{m}_{c \times Q} = 0.48$
     \end{tabular}
     &    
     \begin{tabular}{c}
        ---
     \end{tabular}
     &
     \begin{tabular}{c}
        ---
     \end{tabular}
     &
     \begin{tabular}{c}
         $\tilde{m}_{c}^0 =0.72M_{c}^{\rm sat.}$
     \end{tabular}\\ \hline
 
     \end{tabular}

\end{table}

In this section, we explain the detailed procedures to reproduce the real-space magnetic structure based on the RXS results. First, we performed rocking-curve scans for the magnetic satellites at the $(4, 0, 0) \pm {\bf Q}$ position, and evaluated their exact values of $I_{\pi-\pi'}$ and $I_{\pi-\sigma'}$ by performing the polarization analysis of scattered X-ray. By using Eq. (3) in the main text, we identified the amplitudes of $\tilde{m}_{c}$, $\tilde{m}_{Q}$ and $\tilde{m}_{c \times Q}$ component for ${\bf Q}_{\rm A}=(q_1, 0, 0)$, ${\bf Q}_{\rm A'}=(0, q_2, 0)$, ${\bf Q}_{\rm B}=(q_1/2, q_2/2, 0)$, ${\bf Q}_{\rm B'}=(-q_1/2, q_2/2, 0)$, ${\bf Q}_{\rm C}=(q_1, q_2, 0)$ and ${\bf Q}_{\rm C'}=(-q_1, q_2, 0)$ for Phases II, III, IV and V, as summarized in Supplementary Table I. In Phases II and III, the population ratio of domains $\alpha$ to $\beta$ has been identified from the $I_{\pi-\pi'}$ intensities for ${\bf Q}=(q_1, 0, 0)$ and ${\bf Q}=(0, q_1, 0)$.

On the basis of the experimentally deduced $\tilde{m}_{\alpha}({{\bf Q}_\nu})$ values, the magnetic structure in each magnetic phase is obtained by
\begin{equation}
{\bf m(r)} ={\bf e}_c m_c^0 + \sum_{\alpha, \nu} {\bf e}_\alpha \tilde{m}_{\alpha}({{\bf Q}_\nu}) {\rm sin}({\bf Q}_\nu \cdot {\bf r} + \theta_{\alpha}^{{\bf Q}_\nu}).
\label{spintexture}
\end{equation}
where $\alpha$ represents the directions $c$, $Q$ or $c\times Q$. $m_c^0$ is the uniform magnetization component in the $B \parallel [001]$ direction, which is estimated from the experimental $M$-$B$ profile. While the phase $\theta_{\alpha}^{{\bf Q}_\nu}$ cannot be directly determined from the X-ray scattering experiments, the localized character of Gd$^{3+}$ magnetic moment should result in the spatially uniform $|{\bf m}({\bf r})|$ distribution. By using this constraint, the appropriate $\theta_{\alpha}^{{\bf Q}_\nu}$ values can be uniquely identified. 

\begin{figure}[p]
    \begin{center}
    \includegraphics*[width=15cm]{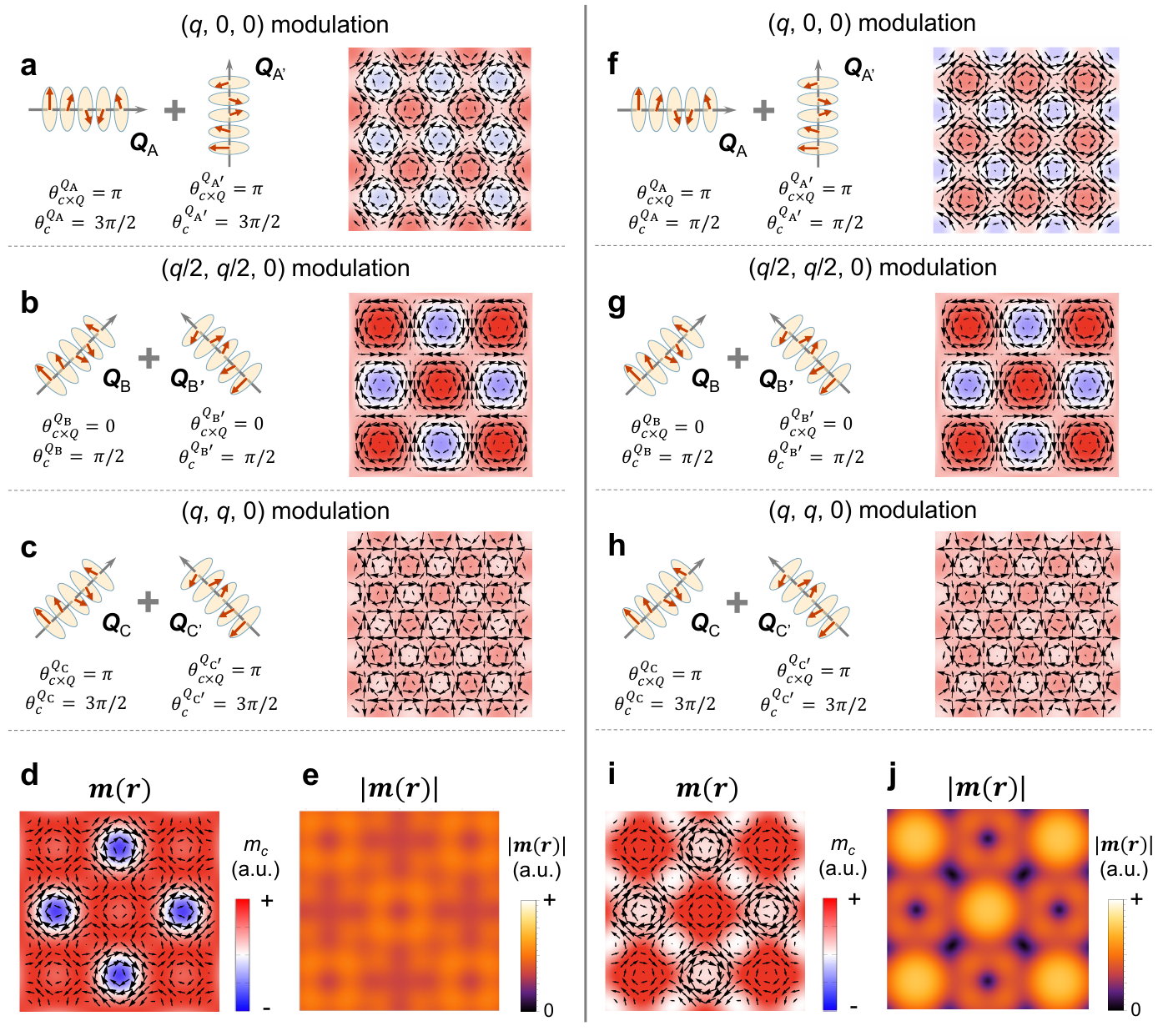}
    \caption{{\bf Reconstruction of magnetic structure m(r) for Phase IV, based on Eq. (S1) and $\tilde{m}_\alpha({{\bf Q}_\nu})$ determined by RXS experiments.} {\bf a}-{\bf c}, Superposition of two screw spin textures characterized by ({\bf a}) ${\bf Q}_{\rm A} = (q, 0, 0)$ and ${\bf Q}_{\rm A'} = (0, q, 0)$, ({\bf b}) ${\bf Q}_{\rm B} = (q/2, q/2, 0)$ and ${\bf Q}_{\rm B'} = (-q/2, q/2, 0)$, and ({\bf c}) ${\bf Q}_{\rm C} = (q, q, 0)$ and ${\bf Q}_{\rm C'} = (q, -q, 0)$. The black arrows and background color represent the in-plane and out-of-plane components of local magnetic moments ${\bf m} ({\bf r})$, respectively. {\bf d}, Superposition of {\bf a}-{\bf c}, which provides the square skyrmion lattice. Here, the relative phase $\theta_{\alpha}^{{\bf Q}_\nu}$ is determined so as to realize the most uniform $|{\bf m} ({\bf r})|$ distribution (See text for the detail). {\bf e}, The real-space distribution of $|{\bf m} ({\bf r})|$ for the spin texture in {\bf d}. The small amount of non-uniform component remaining in the $|{\bf m}({\bf r})|$ profile is associated with the higher-order harmonics neglected in the present model. {\bf f}-{\bf j}, The corresponding ones for the magnetic structure calculated with inappropriate $\theta_{\alpha}^{{\bf Q}_\nu}$, which results in much more inhomogeneous $|{\bf m} ({\bf r})|$ distribution than the one shown in panel {\bf e}, as shown in {\bf j}.}
    \end{center}
\end{figure}

\begin{figure}[p]
\begin{center}
\includegraphics*[width=16cm]{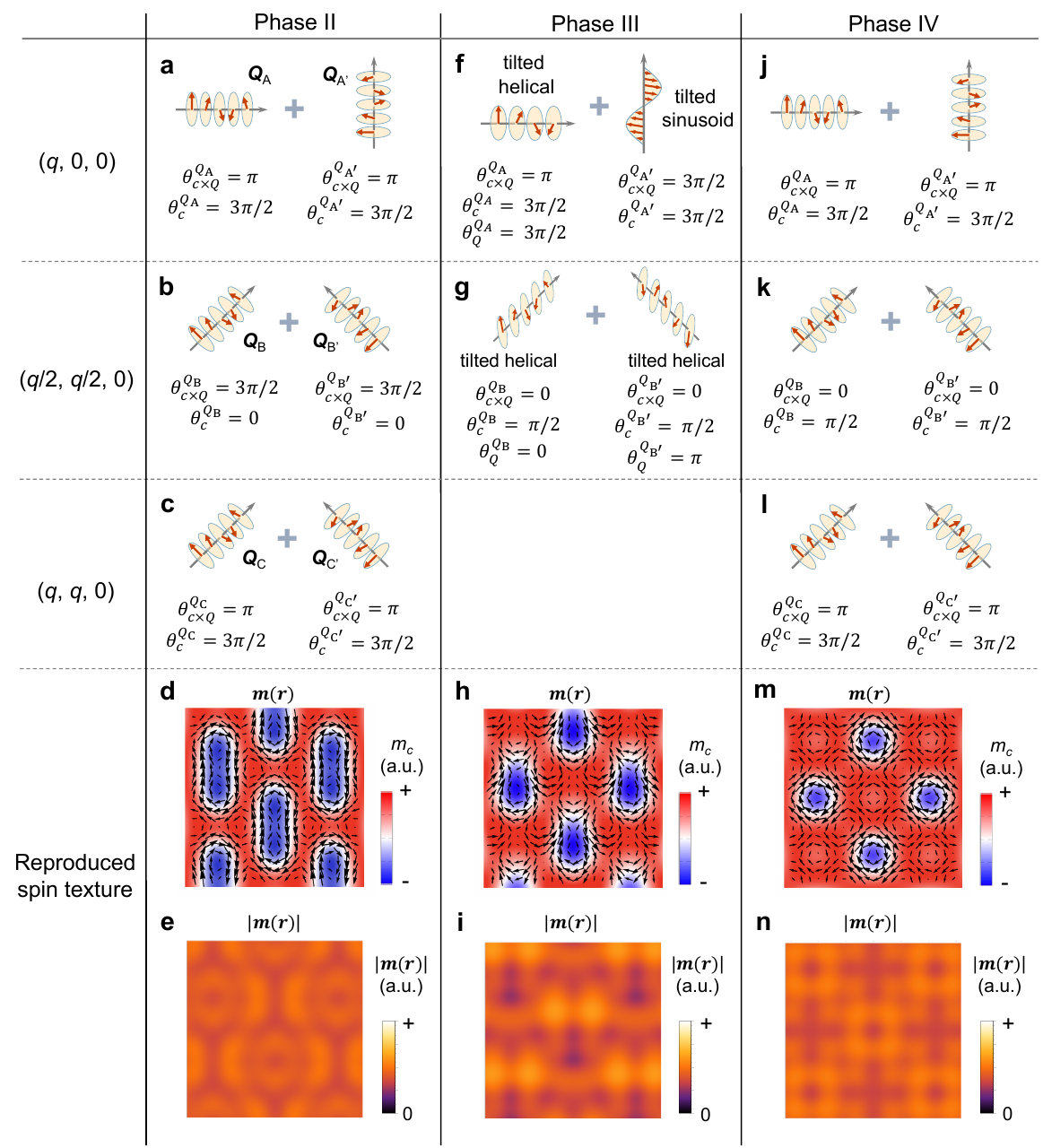}
\caption{{\bf Reconstruction of magnetic structure m(r) for Phases II, III and IV based on RXS results.} {\bf a}-{\bf c}, List of $\theta_{\alpha}^{{\bf Q}_\nu}$ that provides the most uniform $|{\bf m} ({\bf r})|$ distribution for Phase II. {\bf d}, The corresponding magnetic structure ${\bf m} ({\bf r})$ reconstructed based on Eq. (S1) with $\theta_{\alpha}^{{\bf Q}_\nu}$ listed in {\bf a}-{\bf c} and $\tilde{m}_\alpha({{\bf Q}_\nu})$ in Supplementary Table I experimentally deduced from RXS results. {\bf e}, The real-space distribution of local magnetic moment amplitude $|{\bf m} ({\bf r})|$ for the spin texture in {\bf d}. {\bf f}-{\bf i} and {\bf j}-{\bf n}, The corresponding ones for Phases III and IV, respectively.}
\end{center}
\end{figure}

To satisfy this condition, we attempted to deduce the relative phases $\theta_{\alpha}^{{\bf Q}_\nu}$ that provide the most uniform $|{\bf m}({\bf r})|$ distribution. For this purpose, the mean square deviation $\sigma^2_{\rm norm}$ of the local magnetic moment length $|{\bf m} ({\bf r})|$ is defined as 
\begin{equation}
   \label{MSE}
   \sigma^2_{\rm norm} = \frac{1}{A} \int {\rm d}{\bf r} \left ( |{\bf m}({\bf r})| - \overline{|{\bf m}({\bf r})|} \right )^2,
\end{equation}
with $\overline{|{\bf m(r)}|}$ representing the averaged magnetic moment length given by
\begin{equation}
    \overline{|{\bf m} ({\bf r})|} = \frac{1}{A} \int {\rm d}{\bf r} |{\bf m}({\bf r})|.
\end{equation}
Here, the integral is calculated for a magnetic unit cell and $A$ represents its area. For Phases II, III, and IV, we exhaustively investigated all the possible combinations of $\theta_{\alpha}^{{\bf Q}_i}$ (= 0, $\pi$/2, $\pi$, and 3$\pi$/2), and deduced the real-space spin texture that provides the smallest $\sigma^2_{\rm norm}$ value.

Supplementary Figs. 3a-e schematically illustrate the reconstruction process of the real-space spin texture for Phase IV. Supplementary Figs. 3a, b and c indicate the superposition of two screw spin modulations with the wave vectors ${\bf Q}_{\rm A}$ and ${\bf Q}_{\rm A'}$, ${\bf Q}_{\rm B}$ and ${\bf Q}_{\rm B'}$, and ${\bf Q}_{\rm C}$ and ${\bf Q}_{\rm C'}$, respectively. By further taking their superposition based on Eq. (\ref{spintexture}) with the appropriate $\theta_{\alpha}^{{\bf Q}_i}$ values as listed in Supplementary Figs. 3a-c, the square lattice of skyrmions (Supplementary Fig. 3d) with almost uniform $|{\bf m}({\bf r})|$ distribution (Supplementary Fig. 3e) can be obtained. Note that the small amount of non-uniform component remaining in the $|{\bf m}({\bf r})|$ profile is associated with the higher-order harmonics neglected in the present model. On the other hand, when the inappropriate $\theta_{\alpha}^{{\bf Q}_i}$ values are assumed, the $|{\bf m}({\bf r})|$ distribution becomes much more inhomogeneous as shown in Supplementary Figs. 3f-j.

Similar analyses were performed for Phases II, III, and IV. Supplementary Fig. 4 summarizes the reconstructed ${\bf m} ({\bf r})$ and $|{\bf m} ({\bf r})|$ profiles as well as associated $\theta_{\alpha}^{{\bf Q}_i}$ values giving the smallest $\sigma^2_{\rm norm}$. They are characterized by almost uniform $|{\bf m}({\bf r})|$ distribution, and these spin textures are shown in Figs. 4b-d in the main text. Note that the appearance of similar spin textures are also predicted by the theoretical simulation (Fig. 5 in the main text and Supplementary Note VIII), which supports the validity of the present analysis.

\section{RXS measurement of the higher-order magnetic reflections}

In Fig. 2 in the main text, magnetic refection peaks corresponding to ${\bf Q}_{\rm A} \sim (q, 0, 0)$ and ${\bf Q}_{\rm B} \sim (q/2, q/2, 0)$ have been identified. In this section, we further explore the higher-order magnetic reflections corresponding to ${\bf Q}_{\rm C} \sim (q, q, 0)$ and ${\bf Q}_{\rm D} \sim (2q, 0, 0)$. Supplementary Figs. 5g-i summarize the positions of magnetic reflections for Phases II, III and IV, observed in the RXS measurement. We focused on the magnetic satellite reflections around the reciprocal lattice point $(4, 0, 0)$, and performed two kinds of line scans C and D corresponding to the blue and green arrows in Supplementary Fig. 5g, respectively. 

\begin{figure}[b]
    \begin{center}
    \includegraphics*[width=15cm]{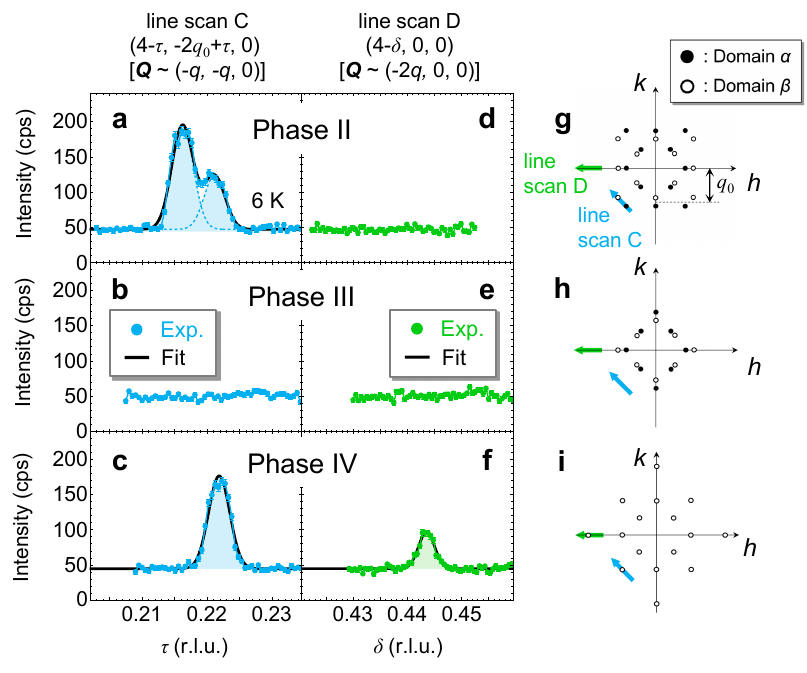}
    \caption{{\bf RXS measurement for the higher-order magnetic satellite peaks.} {\bf a}-{\bf c}, Line profiles of $(4-\tau,\ -2q_0+\tau,\ 0)$ scan (i.e. line scan C) to identify ${\bf Q} \sim (q, q, 0)$ magnetic satellite peaks around the fundamental Bragg spot $(4,0,0)$ for Phases II (0.92 T), III (1.03 T), and IV (1.13 T). The definition of $q_0$ is shown in ({\bf g}). Each experimental data (closed circles) is fitted by one or two Gaussian functions.
    {\bf d}-{\bf f}, Line profiles of $(4-\delta,\ 0,\ 0)$ scan (i.e. line scan D) to identify ${\bf Q} \sim (2q, 0, 0)$ magnetic satellite peaks around the fundamental Bragg spot $(4, 0, 0)$ for Phases II-IV. 
    {\bf g}-{\bf i}, Reciprocal-space distribution of magnetic satellite reflections for Phases II-IV. The directions of line scans C and D are shown in {\bf g}. Closed and open circles in ${\bf g}$ and ${\bf h}$ represent the contributions of magnetic domains $\alpha$ and $\beta$, respectively.}
    \end{center}
\end{figure}

Supplementary Figs. 5a-c indicate the $(4-\tau,\ -2q_0+\tau,\ 0)$ line profiles (i.e. line scan C) in Phases II, III and IV, measured to identify the magnetic reflections corresponding to ${\bf Q}_{\rm C}$. A sharp reflection peak was identified in Phase IV (Supplementary Fig. 5c), which corresponds to the $(4-q, -q, 0)$ reflection. On the other hand, two reflection peaks corresponding to $(4 - q_1, -q_2, 0)$ and $(4 - q_2, - q_1, 0)$ were found in Phase II, where the four-fold symmetry is broken and the coexistence of two equivalent magnetic domains $\alpha$ and $\beta$ leads to the appearance of two peak structure as detailed in Supplementary Note II. In Phase III, no reflection peak was identified in the line scan C (Supplementary Fig. 5b). We have further performed the polarization analysis for these magnetic reflections, and the results are summarized in Supplementary Table 1.

In the similar manner, the $(4-\delta, 0, 0)$ line profiles (i.e. line scan D) were measured to identify the magnetic reflections corresponding to ${\bf Q}_{\rm D}$, and the results for Phases II, III, and IV are plotted in Supplementary Figs. 5d-f. In this case, a magnetic reflection peak was detectable only in the Phase IV, and no peak was identified in Phases II and III. 

Here, the observed scattering intensity for ${\bf Q}_{\rm D}$ (${\bf Q}_{\rm C}$) is at least two (one) orders of magnitude smaller than the fundamental reflection ${\bf Q}_{\rm B}$ (Figs. 2k-m in the main text). Such small harmonic components little affect the magnetic structure. On the basis of the above results, we consider only ${\bf Q}_{\rm A}$, ${\bf Q}_{\rm B}$ and ${\bf Q}_{\rm C}$ for the reconstruction of real-space spin texture in Supplementary Note III.

\section{Meron/anti-meron spin textures in Phase III}

As discussed in the main text, topological charge density $n_{\rm sk} ({\bf r})$ for two-dimensional spin texture ${\bf m}({\bf r})$ is generally defined as 
\begin{equation}
   n_{{\rm sk}} ({\bf r}) = \frac{1}{4\pi} \cdot \left(\frac{\partial {\bf n}}{\partial x} \times \frac{\partial {\bf n}}{\partial y}\right),
\label{nsk}
\end{equation}
where ${\bf n}({\bf r})={\bf m}({\bf r})/|{\bf m}({\bf r})|$ is the unit vector along the local magnetic moment ${\bf m}({\bf r})$. Its spatial integral $N_{\rm sk} = \int n_{\rm sk}({\bf r}){\rm d}x {\rm d}y$ corresponds to the net topological charge, which represents how many times the spin directions wrap a sphere. In general, meron and anti-meron represent the spin texture ${\bf m}({\bf r})$ with $N_{\rm sk} = -1/2$ and $+1/2$, respectively\cite{SkXReviewTokura, Meron1, Meron2}. Supplementary Figs. 6c and d (Supplementary Figs. 6e and f) indicate examples of meron (anti-meron) spin textures, where the spin directions always wrap a half of unit sphere and the in-plane spin component shows vortex-type (anti-vortex-type) arrangement.

\begin{figure}[b]
    \begin{center}
    \includegraphics*[width=16cm]{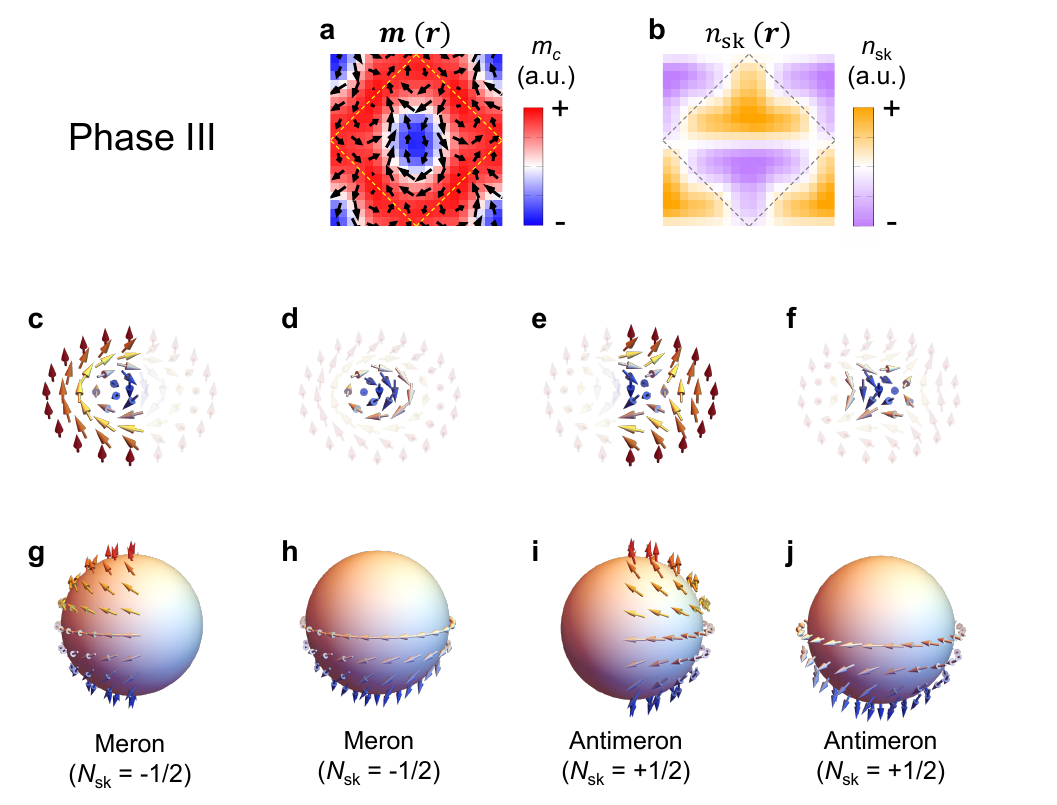}
    \caption{{\bf Topological charge density distribution in Phase III (meron/anti-meron crystal state).} {\bf a}, The magnetic structure ${\bf m} ({\bf r})$ in Phase III obtained by the simulated annealing, which is identical with Fig. 5e in the main text. The black arrows and background color represent the in-plane and out-of-plane component of local magnetic moment ${\bf m}({\bf r})$, respectively. {\bf b}, Corresponding topological charge density distribution $n_{\rm sk} ({\bf r})$ in Phase III. The background color represents the amplitude of $n_{\rm sk}$ calculated based on Eq. (\ref{nsk}). Magnetic unit cell is indicated with dotted lines in {\bf a} and {\bf b}. {\bf c}-{\bf f}, Examples of meron and anti-meron spin textures characterized by $N_{\rm sk} = -1/2$ and  $N_{\rm sk} = +1/2$, respectively. {\bf g}-{\bf j}, Their projections onto a unit sphere.}
    \end{center}
\end{figure}

On the basis of the above framework, we analyze the spin texture in Phase III. Supplementary Fig. 6a indicates the magnetic structure ${\bf m}({\bf r})$ in Phase III, obtained by the theoretical simulation in the main text. This spin texture can be considered as the combination of Supplementary Figs. 6c and e, i.e. a pair of meron and anti-meron. To confirm the validity of this picture, the corresponding spatial distribution of topological charge density $n_{\rm sk} ({\bf r})$ is plotted in Supplementary Fig. 6b. Here, the magnetic unit cell is indicated by the dashed square, and its upper and lower half are characterized by the positive and negative sign of local topological charge, respectively. It supports that the spin texture in Phase III can be regarded as the meron/anti-meron lattice.

In terms of the $n_{\rm sk} ({\bf r})$ distribution, the presently identified meron/anti-meron pair is identical with so-called type-II magnetic bubble. This object can behave like a particle (i.e. "molecule" of meron and anti-meron) as a whole\cite{TypeIIBubble}, since the opposite sign of out-of-plane spin component between the core and edge region causes sizable energy barrier for its elimination. Previously, type-II magnetic bubbles were studied in ferromagnetic thin films with dipolar/exchange interactions and perpendicular anisotropy, while they usually appear as a meta-stable state and their typical diameter was in order of several micrometers\cite{BubbleTextbook}. In contrast, our present results revealed that the competition of itinerant-electron-mediated interactions can stabilize a meron/anti-meron pair even in the equilibrium ground state, whose diameter (2.7 nm) is three orders of magnitude smaller than the typical magnetic bubbles. 

Note that the vortex lattice spin texture in Phase V (Fig. 5g in the main text) was often termed as "meron/anti-meron-like" state in the previous literature\cite{GdRu2Si2_Full}. In this state, however, the local spin texture can wrap only less than a half of sphere (i.e. $|N_{\rm sk}| < 1/2$) due to the uniform out-of-plane spin component, and Phase V cannot be considered as a genuine meron/anti-meron lattice state.

\section{Photon Energy dependence of magnetic X-ray scattering}

\begin{figure}[t]
    \begin{center}
    \includegraphics*[width=9cm]{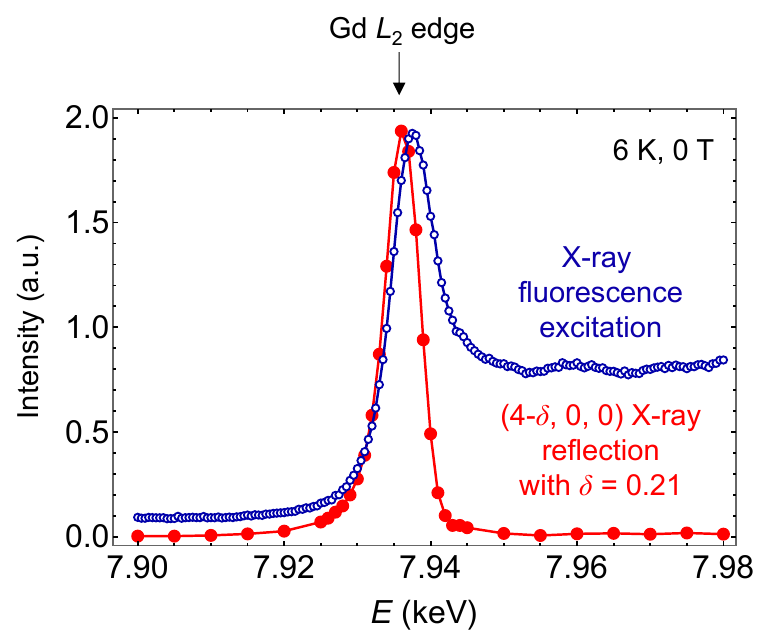}
    \caption{{\bf Photon-energy dependence of the X-ray reflection and fluorescence excitation.} Closed red symbols show the scattering intensity of the ($4-\delta$, 0, 0) magnetic X-ray reflection with $\delta = 0.21$ measured at 6 K and 0 T. The maximum value at $E \sim 7.935$ keV corresponds to the Gd $L_2$ absorption edge. Blue open symbols represent the X-ray fluorescence excitation.}
    \end{center}
\end{figure}

Supplementary Fig. 7 indicates the photon-energy dependence of the X-ray fluorescence excitation and the magnetic scattering intensity for the ($4-\delta$, 0, 0) reflection with $\delta = 0.21$ measured at $B=0$. A resonance peak was observed at $E \sim 7.935$ keV, which corresponds the Gd $L_2$ absorption edge representing the transition from $2p$ orbital to $5d$ orbital of Gd. This result proves that the observed RXS results indeed reflect the magnetic properties of the Gd sites.

\section{Neutron scattering profiles in Phases I, II, III, IV and V}

\begin{figure}[p]
   \begin{center}
   \includegraphics*[width=14.5cm]{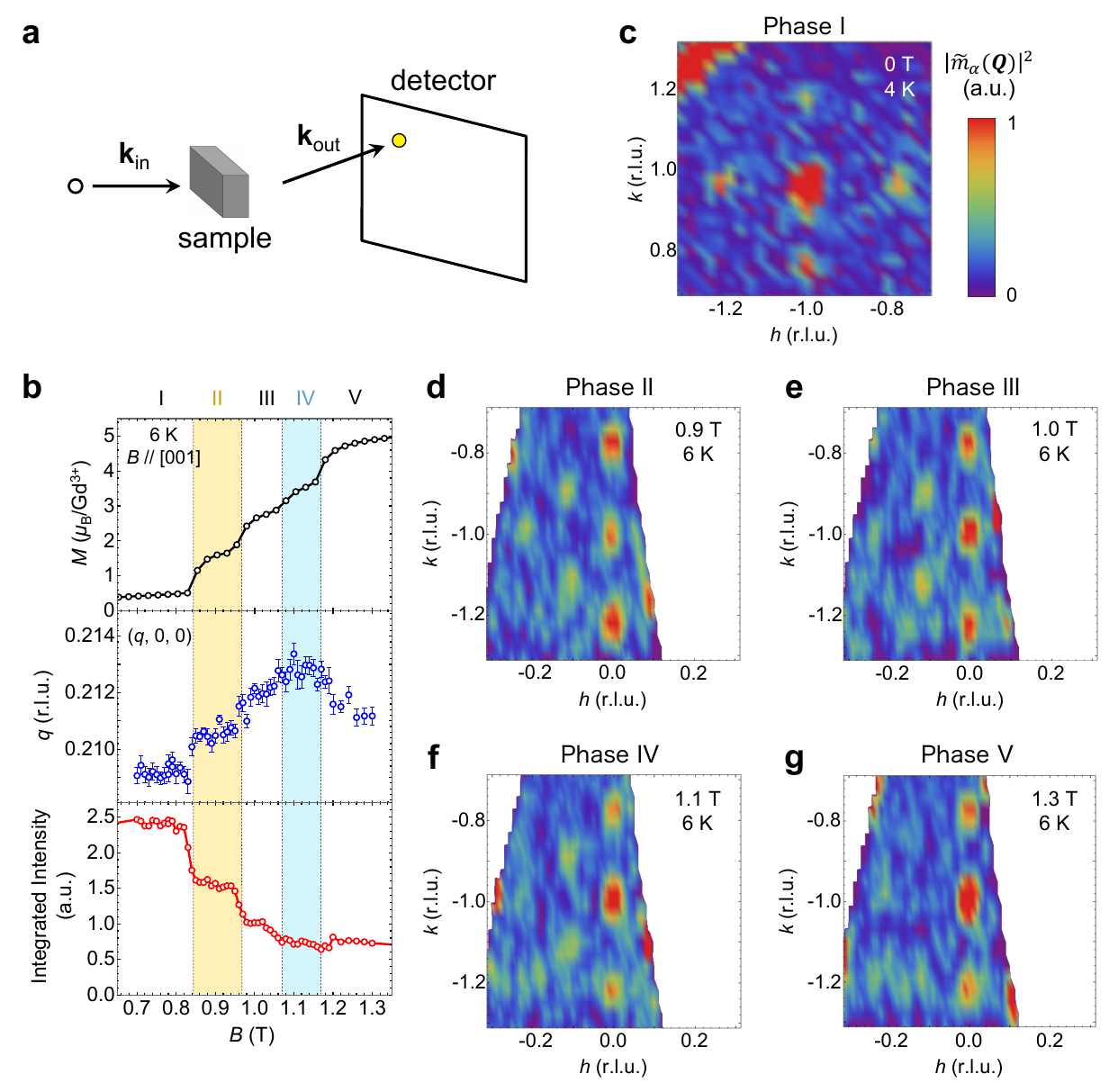}
   \caption{{\bf Neutron scattering experiments in Phases I, II, III, IV and V for GdRu$_2$Ge$_2$.} {\bf a}, Schematic illustration of the setup for the neutron scattering measurements. ${\bf k}_{\rm in}$ and ${\bf k}_{\rm out}$ represent the propagation vectors of incident and scattered neutron beams, respectively. {\bf b}, Magnetic-field dependence of magnetization $M$, wavenumber $q$, and integrated intensity for the $(q, 0, 0)$ magnetic reflection. $M$ in the upper panel of {\bf b} is measured by using a SQUID magnetometer. In the middle panel of {\bf b}, the error bars are tentatively plotted based on the standard errors for the Gaussian fitting of experimental data. Note that the actual error bars would be much larger, because of the small unintended misalignment of the crystal orientation that is inevitable for the high-energy neutron scattering experiments with limited wave-number resolution. {\bf c}, Reciprocal-space distribution of neutron scattering intensity around the $(-1, 1, 0) \pm {\bf Q}$ position measured for Phase I ($B=0$). {\bf d}-{\bf g}, The corresponding data around the $(0, -1, 0) \pm {\bf Q}$ position for Phases II, III, IV and V, measured at $B= 0.9$, 1.0, 1.1 and 1.3 T, respectively.}
   \end{center}
\end{figure}

In this section, we discuss the results of neutron scattering experiments for GdRu$_2$Ge$_2$. To reduce the large neutron absorption of Gd, the energy of the incident neutron beam was tuned to 153.5 meV by using a high-resolution Fermi chopper, as detailed in Methods section.

Supplementary Fig. 8b indicates the magnetic-field dependence of magnetization $M$, as well as the wavenumber $q$ and integrated intensity of the $(q,0,0)$ magnetic reflection, measured for $B \parallel [001]$ at 6 K. The magnetic phase transitions characterized by step-like magnetization anomalies are accompanied by an abrupt change of $q$-value and/or integrated intensity. These behaviors are in accord with the ones observed by the resonant X-ray scattering experiments in Figs. 2b-d in the main text.

In Supplementary Figs. 8c-g, reciprocal-space distribution of neutron scattering intensity for the $(hk0)$ plane measured in Phases I, II, III, IV, and V are plotted. These data revealed the magnetic satellites at $(-1, 1, 0) \pm {\bf Q}$ or $(0, -1, 0) \pm {\bf Q}$ positions. Here, magnetic reflections corresponding to ${\bf Q_{\rm A}} = (q, 0, 0)$ are always observed, and additional peaks corresponding to ${\bf Q_{\rm B}} \sim (q/2, q/2, 0)$ appear in Phases II, III, and IV. These results are consistent with the scattering patterns identified by the RXS experiments (Fig. 2 in the main text).

Note that the imbalance of scattering intensity between the $(\pm q, -1, 0)$ and $(0, -1 \pm q, 0)$ peaks in Supplementary Figs. 8d-g arises from the selection rule of neutron scattering. In general, the neutron scattering intensity $I \propto  |\tilde{{\bf m}}^{\perp}({\bf Q})|^2$ reflects the modulated spin component normal to the scattering vector ($\tilde{{\bf m}}^{\perp}({\bf Q})$). In case of the $(0, -1, 0) \pm {\bf Q}$ reflection, the scatting vector is almost parallel to the [010] direction, and therefore the scattering intensity should mainly reflect the [100] and [001] component of $\tilde{{\bf m}}({\bf Q})$. According to Supplementary Table 1, the [100] component of $\tilde{{\bf m}}({\bf Q})$ for ${\bf Q_{\rm A}} = (q, 0, 0)$ is much smaller than the one for ${\bf Q_{\rm A'}} = (0, q, 0)$, which explains the stronger scattering intensity in the $(0, -1\pm q, 0)$ reflections. While both ${\bf Q_{\rm A}}$ and ${\bf Q_{\rm A'}}$ also possess the common [001] component of $\tilde{{\bf m}}({\bf Q})$, its amplitude is gradually suppressed as $B \parallel [001]$ increases. It causes the larger imbalance of scattering intensity between ${\bf Q_{\rm A}}$ and ${\bf Q_{\rm A'}}$ for larger $B$, in consistent with the experimental observations.

Here, the employment of high-energy neutron compromises the wave-number resolution in the reciprocal space, which leads to much broader magnetic reflection peaks than the resonant X-ray scattering experiments. As a result, the two distinctive magnetic satellite peaks at the $(q_1, 0, 0)$ and $(q_2, 0, 0)$ positions for Phases II and III (Fig. 2 in the main text) are observed as a broad single peak in the present case. In Supplementary Fig. 8b, the error bar for the $q$-value is tentatively plotted based on the asymptotic standard errors of the peak positions in the least-squares fitting analysis for the observed neutron diffraction profiles using Gaussian functions. Nevertheless, the actual error bar would be much larger due to possible imperfections in aligning the single-crystal sample with respect to the incident neutron beam, which is inevitable for the high-energy neutron small-angle scattering experiments with limited wavenumber resolution. This effect is particularly large for the ($q$, 0, 0) magnetic reflection considered in Supplementary Fig. 8b, characterized by the small neutron scattering angle ($\sim 1.8 ^\circ$). The above factors cause the slight inconsistency of $q$-value between Supplementary Fig. 8b (i.e. neutron scattering) and Fig. 2c in the main text (i.e. resonant X-ray scattering). Considering the compromised wave-number resolution in the former neutron data, the latter X-ray data should be more reliable.

\section{Theoretical simulation of stable magnetic structures for $B \parallel [001]$}

To understand the microscopic origin of nontrivial magnetic orders and successive magnetic phase transitions in GdRu$_2$Ge$_2$, we performed the simulated annealing for the two-dimensional square lattice system based on the effective magnetic Hamiltonian derived from the Kondo lattice model\cite{HayamiModel1, HayamiModel2, Hayami_JPSJ} given by
\begin{equation}
\mathcal{H} = -2J \sum_{\nu, \alpha, \beta} \Gamma^{\alpha \beta}_{{\bf Q}_\nu} {\tilde m}_{\alpha}({{\bf Q}_\nu}) {\tilde m}_{\beta}({{\bf Q}_\nu}) - \sum_i {\bf B} \cdot {\bf m} ({\bf r}_i)
\label{Hamiltonian}
\end{equation}
with $\alpha,\beta = [100], [010], [001]$ and ${\bf Q}_{\nu}= {\bf Q}_{\rm A}, {\bf Q}_{\rm A'}, {\bf Q}_{\rm B}, {\bf Q}_{\rm B'}$. $\tilde{{\bf m}} ({\bf Q}_\nu)$ is the Fourier transform of the real-space distribution of classical localized spin ${\bf m} ({\bf r}_i)$, whose amplitude is fixed at $|{\bf m} ({\bf r}_i)|=1$. Here, the first and second terms represent RKKY interaction and Zeeman energy, respectively. We suppose the situation shown in Supplementary Figs. 9a and b, where the bare susceptibility $\chi ({\bf Q})$ of itinerant electron shows the first maxima at ${\bf Q}_{\rm A} =(q, 0)$ and ${\bf Q}_{\rm A'} =(0,q)$ with $q = 2\pi/5$, and the relatively large values at ${\bf Q}_{\rm B} =(q/2, q/2)$ and ${\bf Q}_{\rm B'} =(-q/2, q/2)$ satisfying the relation ${\bf Q}_{\rm A} = {\bf Q}_{\rm B} - {\bf Q}_{\rm B'}$ and ${\bf Q}_{\rm A'} = {\bf Q}_{\rm B} + {\bf Q}_{\rm B'}$. Such $\chi({\bf Q})$ distribution can be naturally realized by considering the nesting of the Fermi surfaces, for example\cite{HayamiModel1, Hayami_JPSJ}. For these ordering vectors, we adjusted the value of interaction tensors $\Gamma^{\alpha \beta}_{{\bf Q}_\nu}$ to satisfy the tetragonal lattice symmetry, weak easy-axis anisotropy, and the relation $\chi({\bf Q}_{\rm A}) = \chi({\bf Q}_{\rm A'}) > \chi({\bf Q}_{\rm B}) = \chi({\bf Q}_{\rm B'})$; $J=1$, $\kappa=0.86$, $\gamma_1=0.9$, $\gamma_2=0.91675$, and $\gamma_3=0.0725$ for 
$\Gamma^{xx}_{\mathbf{Q}_{\rm A}}=\Gamma^{yy}_{\mathbf{Q}_{\rm A'}}=\gamma_1\gamma_2$, 
$\Gamma^{yy}_{\mathbf{Q}_{\rm A}}=\Gamma^{xx}_{\mathbf{Q}_{\rm A'}}=\gamma_1$, 
$\Gamma^{zz}_{\mathbf{Q}_{\rm A}}=\Gamma^{zz}_{\mathbf{Q}_{\rm A'}}=1$, 
$\Gamma^{xx}_{\mathbf{Q}_{\rm B}}=\Gamma^{yy}_{\mathbf{Q}_{\rm B}}=\Gamma^{xx}_{\mathbf{Q}_{\rm B'}}=\Gamma^{yy}_{\mathbf{Q}_{\rm B'}}=\kappa \gamma_1$,
$-\Gamma^{xy}_{\mathbf{Q}_{\rm B}}=-\Gamma^{yx}_{\mathbf{Q}_{\rm B}}=\Gamma^{xy}_{\mathbf{Q}_{\rm B'}}=\Gamma^{yx}_{\mathbf{Q}_{\rm B'}}=\kappa \gamma_3$,
for $\Gamma^{zz}_{\mathbf{Q}_{\rm B}}=\Gamma^{zz}_{\mathbf{Q}_{\rm B'}}=\kappa$.

\begin{figure}[p]
   \begin{center}
   \includegraphics*[width=15cm]{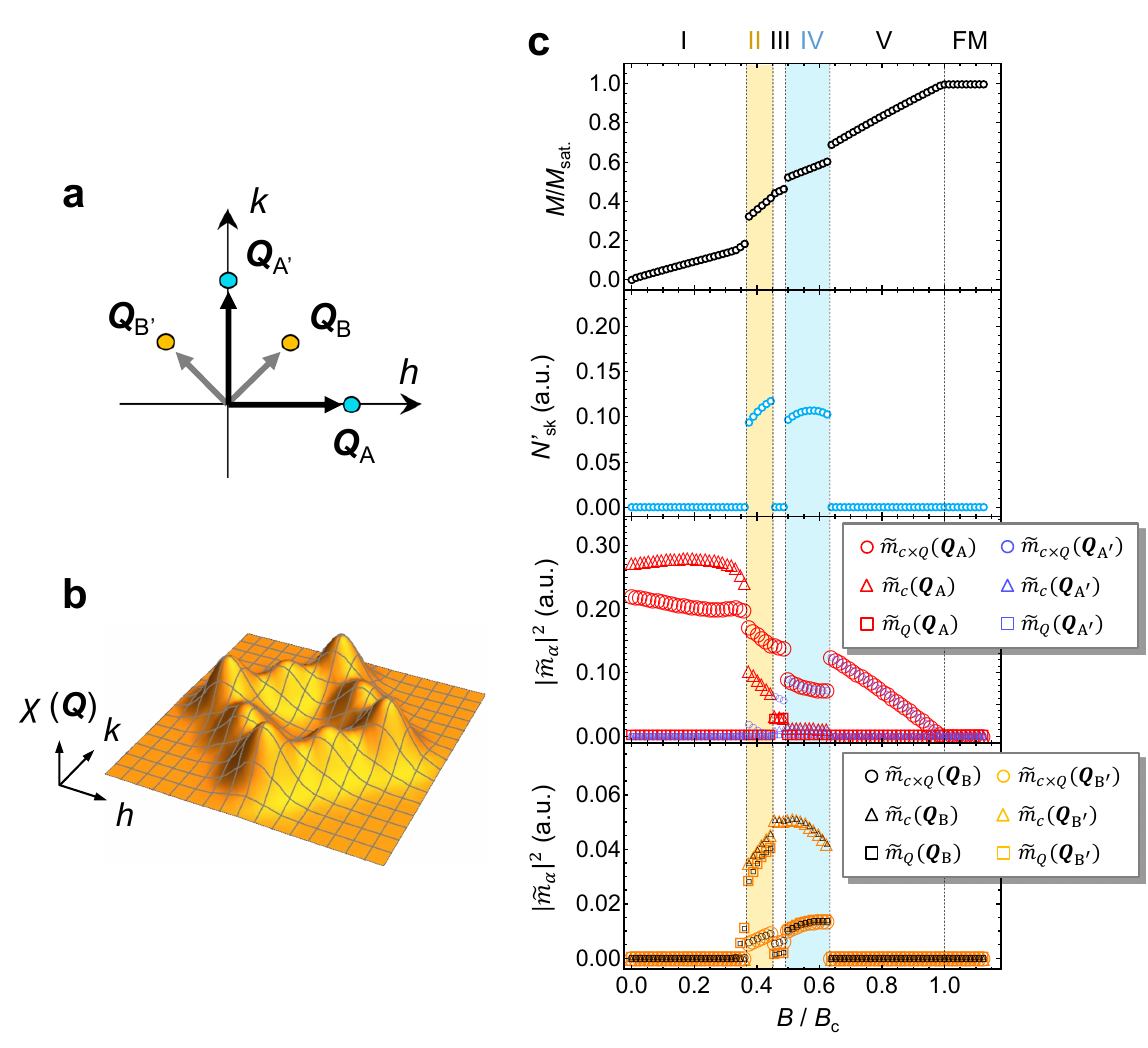}
   \caption{{\bf Theoretically simulated magnetic-field dependence of spin textures.} {\bf a},{\bf b}, Schematic illustration of bare susceptibility $\chi ({\bf Q})$ distribution considered in the effective magnetic Hamiltonian Eq. (\ref{Hamiltonian}), which assumes the largest peaks at ${\bf Q}_{\rm A} = (q, 0, 0)$ and ${\bf Q}_{\rm A'} = (0, q, 0)$ as well as the relatively large value at ${\bf Q}_{\rm B} = (q/2, q/2, 0)$ and ${\bf Q}_{\rm B'} = (-q/2, q/2, 0)$. {\bf c}, Magnetic-field dependence of magnetization $M$, scalar spin chirality $N'_{\rm sk}$ (defined in Methods section) and modulated spin component $m_{\alpha}({{\bf Q}_\nu})$, theoretically calculated based on magnetic Hamiltonian in Eq. (\ref{Hamiltonian}) with magnetic field $B$ applied normal to the square lattice. $M_{\rm sat.}$ and $B_{\rm c}$ represents the saturated magnetization and the critical magnetic field to obtain fully polarized ferromagnetic (FM) state, respectively. Phases II and IV are highlighted by yellow and blue shadows.}
   \end{center}
\end{figure}

Supplementary Fig. 9c indicates the magnetic-field dependence of magnetization $M$, scalar spin chirality $N'_{\rm sk}$, and various ${\tilde m}_{\alpha} ({{\bf Q}_\nu})$ components for $B \parallel [001]$ theoretically calculated based on Eq. (\ref{Hamiltonian}). The definition of scalar spin chirality $N'_{\rm sk}$ is provided in Methods section, which becomes non-zero for $N_{\rm sk} \neq 0$. The simulation results predict the appearance of successive magnetic phase transitions (Phase I $\rightarrow$ II $\rightarrow$ III $\rightarrow$ IV $\rightarrow$ V $\rightarrow$ FM). Here, the spin components modulated with ${\bf Q}_{\rm B}$ and ${\bf Q}_{\rm B'}$ appear only in the Phases II, III, and IV, and the emergence of finite scalar spin chirality is confirmed for the Phases II and IV. Theoretically obtained spin texture ${\bf m} ({\bf r})$ for each magnetic phase is summarized in Figs. 5c-g in the main text, which well reproduces the experimentally deduced ones in Figs. 4a-e in the main text. The overall good agreement between the theoretical and experimental results supports the validity of our magnetic structure analysis in Supplementary Note III, and suggests that the magnetism in GdRu$_2$Ge$_2$ is well captured by Eq. (\ref{Hamiltonian}).

Note that a non-centrosymmetric chiral magnet Co$_8$Zn$_9$Mn$_3$ was previously reported to exhibit a transition between the square meron lattice and hexagonal skyrmion lattice states\cite{Meron1}. Nevertheless, the skyrmion formation in this compound requires DM interaction, and the mechanism of observed meron-skyrmion transition has not been fully clarified yet. On the other hand, our present results demonstrate that more intricate manner of multi-step topological magnetic transitions (among elliptic skyrmion, meron/anti-meron pair and circular skyrmion) can be realized even in a centrosymmetric compound without DM interaction. Its skyrmion diameter is almost two orders of magnitude smaller than the traditional DM-based mechanism. Our analysis revealed that a novel mechanism, i.e. the competition of RKKY interactions at inequivalent wave vectors, is the key for the observed multi-step topological transitions, which will contribute to the better understanding and further exploration of non-trivial topological magnetic quasi-particles in centrosymmetric systems.

\section{Microscopic origin of different magnetic phase diagrams between GdRu$_2$Ge$_2$ and GdRu$_2$Si$_2$}

In this section, to better understand the origin of the multi-step topological phase transitions in GdRu$_2$Ge$_2$, we analyze the key difference between the previously studied GdRu$_2$Si$_2$ and the present GdRu$_2$Ge$_2$. In case of GdRu$_2$Si$_2$, the $M$-$B$ profile for $B \parallel [001]$ shows only a single intermediate step, where the square lattice of circular skyrmion is realized\cite{GdRu2Si2}. On the other hand, GdRu$_2$Ge$_2$ hosts three intermediate magnetization steps, which correspond to the square lattice states of elliptic skyrmion, meron/anti-meron pair, and circular skyrmion.

\begin{figure}[p]
    \begin{center}
    \includegraphics*[width=16cm]{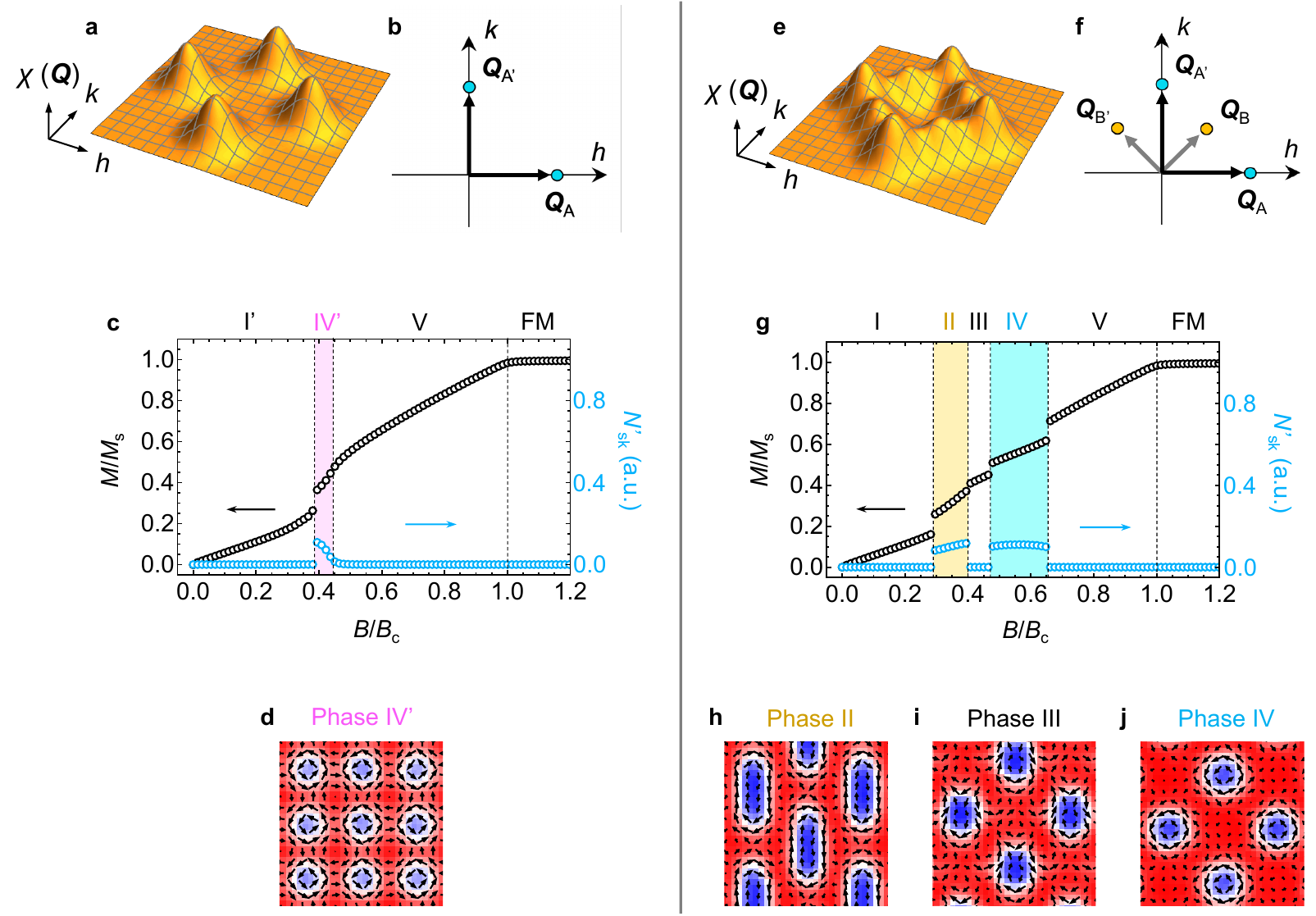}
    \caption{{\bf Theoretical simulation with and without the competition of RKKY interactions at inequivalent wave vectors.} {\bf a},{\bf b}, Schematic illustration of bare susceptibility $\chi ({\bf Q})$ distribution with the peak structures at ${\bf Q}_{\rm A} = (q, 0)$ and ${\bf Q}_{\rm A'} = (0, q)$. {\bf c}, Magnetic-field dependence of magnetization $M$ and scalar spin chirality $N'_{\rm sk}$ (defined in Methods section), theoretically calculated based on magnetic Hamiltonian in Eq. (\ref{Supple_Hamiltonian}) and $\chi ({\bf Q})$ distribution in {\bf a}. Here, the value of interaction tensor $\Gamma^{\alpha \beta}_{{\bf Q}_{\rm \nu}}$ for ${{\bf Q}_{\rm \nu}} = {\bf Q}_{\rm A}$ is the same as the ones used for Eq. (6) in the main text or Eq. (\ref{Hamiltonian}) in Supplementary Note VIII, while the one for ${{\bf Q}_{\rm \nu}} = {\bf Q}_{\rm B}$ is zero. {\bf d}, Theoretically simulated magnetic structure ${\bf m} ({\bf r})$ in Phases IV', obtained at $B = 0.39B_c$. {\bf e}-{\bf j}, The corresponding ones for the $\chi ({\bf Q})$ distribution with the largest peaks at ${\bf Q}_{\rm A}$ and ${\bf Q}_{\rm A'}$ as well as relatively large value at ${\bf Q}_{\rm B} = (q/2, q/2)$ and ${\bf Q}_{\rm B'} = (-q/2, q/2)$. In this case, the values of interaction tensor $\Gamma^{\alpha \beta}_{{\bf Q}_{\rm \nu}}$ for ${{\bf Q}_{\rm \nu}} = {\bf Q}_{\rm A}$ and ${{\bf Q}_{\rm \nu}} = {\bf Q}_{\rm B}$ are the same as the ones used for Eq. (6) in the main text or Eq. (\ref{Hamiltonian}) in Supplementary Note VIII. ${\bf h}, {\bf i}$ and ${\bf j}$ indicate  theoretically simulated magnetic structures in Phases II, III, and IV, obtained at $B = 0.33B_c$, $B = 0.45B_c$, and $B = 0.56B_c$, respectively. Note that $K=0.1$ is assumed throughout this figure, in contrast to Supplementary Fig. 9 that assumes $K=0$.}
    \end{center}
\end{figure}

To understand the magnetism in these systems, we performed the simulated annealing for the two-dimensional square lattice system based on the effective magnetic Hamiltonian given by
\begin{equation}
    \mathcal{H} = -2\sum_{\nu}\left\{J \lambda_{{\bf Q}_\nu} 
    + \frac{K}{N}\lambda_{{\bf Q}_\nu}^2 \right\}
    - \sum_i {\bf B} \cdot {\bf m} ({\bf r}_i)
    \label{Supple_Hamiltonian}
\end{equation}
with
\begin{equation}
    \lambda_{{\bf Q}_\nu} = \sum_{\alpha, \beta} \Gamma^{\alpha \beta}_{{\bf Q}_\nu} {\tilde m}_{\alpha}({{\bf Q}_\nu}) {\tilde m}_{\beta}({{\bf Q}_\nu}),
    \label{Supple_Hamiltonian_lumbda}
\end{equation}
which was originally derived from Kondo lattice model in Ref. \cite{HayamiModel1, HayamiModel2, Hayami_JPSJ}. Here, $J(\equiv 1)$ and $K$ represent the amplitude of RKKY and four-spin interactions, respectively. The definition of each symbol is common with Eq. (\ref{Hamiltonian}). In the following, we assume $K = 0.1$, and explore how the magnetic interactions at inequivalent wave vectors affect the magnetic phase diagram.

First, we consider the magnetic interactions at ${\bf Q}_{\rm A} =(q, 0)$ and ${\bf Q}_{\rm A'} =(0,q)$ only. It corresponds to the magnetic susceptibility distribution $\chi ({\bf Q})$ as shown in Supplementary Figs. 10a and b, which is characterized by $\chi ({\bf Q})$ peaks at $\pm {{\bf Q}_{\rm A}}$ and $\pm {{\bf Q}_{\rm A'}}$. Supplementary Fig. 10c indicates the magnetic field dependence of magnetization $M$ and scalar spin chirality $N'_{\rm sk}$ calculated by the simulated annealing based on Eq. (\ref{Supple_Hamiltonian}). The magnetization profile shows only a single intermediate step (Phase IV'), and it represents the square lattice of circular skyrmion (Supplementary Fig. 10d). This model well reproduces the magnetic phase diagram in GdRu$_2$Si$_2$\cite{GdRu2Si2_Full}.

Next, we consider the magnetic interactions at ${\bf Q}_{\rm B} =(q/2, q/2)$ and ${\bf Q}_{\rm B'} =(q/2, -q/2)$, in addition to the ones at ${\bf Q}_{\rm A}$ and ${\bf Q}_{\rm A'}$. It corresponds to the magnetic susceptibility distribution $\chi ({\bf Q})$ as shown in Supplementary Figs. 10e and f, where the largest peaks appear at  $\pm {\bf Q}_{\rm A}$ and $\pm {\bf Q}_{\rm A'}$ and relatively large value appear at $\pm {\bf Q}_{\rm B}$ and $\pm {\bf Q}_{\rm B'}$. The simulated magnetic field dependence of $M$ and $N'_{\rm sk}$ is plotted in Supplementary Fig. 10g. In this case, there appear three intermediate magnetization steps (Phases II, III and IV), which represent the square lattice of elliptic skyrmion, meron/anti-meron pair, and circular skyrmion, respectively (Supplementary Figs. 10h-j). It well reproduces the experimental behavior of GdRu$_2$Ge$_2$. When $K=0$ is assumed, this model becomes identical with Eq. (\ref{Hamiltonian}) in Supplementary Note VIII. Even in the latter case without four-spin interaction $K$, the same phase transition process (Phases II $\rightarrow$ III $\rightarrow$ IV) is obtained as shown in Supplementary Fig. 9.

The above analyses indicate that the competition of RKKY interactions at inequivalent wave vectors  ${\bf Q}_{\rm A}$ and ${\bf Q}_{\rm B}$ is the key for the appearance of multi-step topological phase transitions in GdRu$_2$Ge$_2$, and the four-spin interaction $K$ is not essential. This is distinct from previous theoretical models proposed for GdRu$_2$Si$_2$, where only the interaction at ${\bf Q}_{\rm A}$ was considered. The present results demonstrate that the competition of RKKY interactions at inequivalent wave vectors ${\bf Q}_{\rm A} = (q, 0)$ and ${\bf Q}_{\rm B} = (q/2, q/2)$ (typically induced by the nesting of Fermi-surfaces along multiple directions\cite{HayamiModel1, Hayami_JPSJ}) can be a novel promising route to realize a rich variety of topological magnetic quasi-particles in a single compound, which would be a good guideline for further exploration of non-trivial swirling spin textures.

\section{Analysis of Hall resistivity profiles}

In this section, we discuss the microscopic origin of Hall signals in GdRu$_2$Ge$_2$. Supplementary Figs. 11a and b indicate the magnetic-field dependence of magnetization $M$ and Hall resistivity $\rho_{yx}$ measured for $B \parallel [001]$ at various temperatures. At 2 K, a series of step-like anomalies are observed in the magnetization profile, which represent the magnetic phase transitions among the Phases I, II, III, IV, and V. Correspondingly, Hall resistivity profile also shows peak-like enhancement in Phases II and IV (i.e. shaded regions in Supplementary Fig. 11b). These anomalies become weaker at higher temperature, and the $\rho_{yx}$ peak structures disappear above 26 K. On the basis of these data, the $B$-$T$ magnetic phase diagram is summarized in Figs. 1h and i in the main text.

\begin{figure}[b]
   \begin{center}
   \includegraphics*[width=17cm]{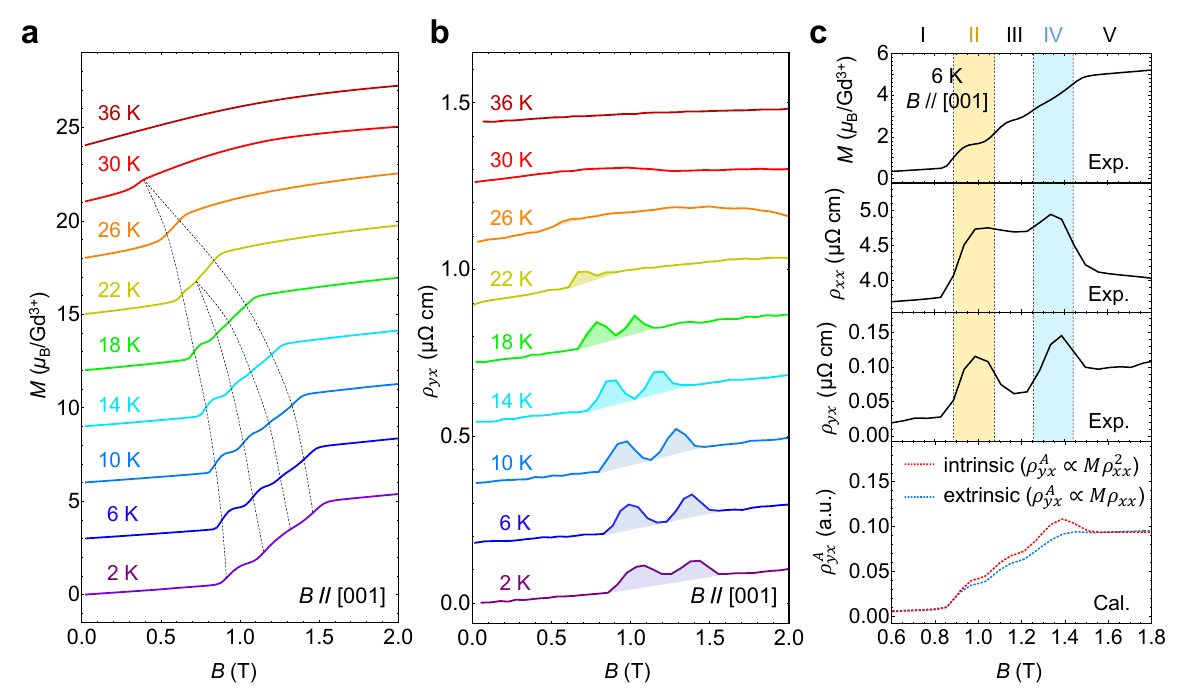}
   \caption{{\bf Magnetic and electrical transport properties of GdRu$_2$Ge$_2$ measured at various temperature.} {\bf a,b}, Magnetic-field dependence of magnetization $M$ and Hall resistivity $\rho_{yx}$ measured for $B \parallel [001]$ at various temperatures. Each data is arbitrarily shifted along the vertical direction for clarity. Dotted lines indicate the temperature development of magnetic phase boundaries. Two peak structures in $\rho_{yx}$, probably representing the topological Hall term $\rho^{\rm T}_{yx}$, are highlighted by the shadows. {\bf c}, Magnetic-field dependence of $M$, longitudinal resistivity $\rho_{xx}$, and $\rho_{yx}$ measured at 6 K. The expected $B$-dependence of anomalous Hall term $\rho^{\rm A}_{yx}$, calculated based on $\rho^{\rm A}_{yx} \propto M \rho_{xx}^2$ and $\rho^{\rm A}_{yx} \propto M \rho_{xx}$ assuming the intrinsic and extrinsic mechanism, respectively, are also plotted in the bottom panel.}
   \end{center}
\end{figure}

As discussed in the main text, Hall resistivity $\rho_{yx}$ is generally described as
\begin{equation}
\rho_{yx} = \rho_{yx}^{{\rm N}}+\rho_{yx}^{{\rm A}}+\rho_{yx}^{{\rm T}} = R_0 B + R_s M + P R_0 B_{\rm em}\mathrm{,}
\end{equation}
where $\rho_{yx}^{{\rm N}} \propto B$ and $\rho_{yx}^{{\rm A}} \propto M$ represent the normal and anomalous Hall terms, respectively ($R_0$ and $R_s$ are the coefficients for the respective terms)\cite{SkXReviewTokura, THE, RMP_AHE}. According to recent theoretical studies, anomalous Hall effect can originate from the intrinsic and extrinsic mechanisms, which are characterized by the relations $R_s \propto \rho_{xx}^2$ and $R_s \propto \rho_{xx}$, respectively\cite{RMP_AHE}. The third term $\rho_{yx}^{{\rm T}}$ represents the topological Hall effect, which scales with the emergent magnetic-field $B_{\rm em} = (h/e) n_{\rm sk}$ in proportion to skyrmion density $n_{\rm sk}$\cite{SkXReviewTokura, RMP_AHE}. $P$ represents the spin polarization ratio of conduction electrons. 

Supplementary Fig. 11c indicates the magnetic-field dependence of magnetization $M$, longitudinal resistivity $\rho_{xx}$, and Hall resistivity $\rho_{yx}$ measured at 6 K for $B \parallel [001]$. In the bottom panel of Supplementary Fig. 11c, we also plotted the calculated profiles of $\rho_{yx}^{\rm A} \propto M \rho_{xx}^2$ and $\rho_{yx}^{\rm A} \propto M \rho_{xx}$ to estimate the expected $B$-dependence of anomalous Hall term $\rho_{yx}^{\rm A}$ associated with the intrinsic and extrinsic mechanisms, respectively. In both cases, the experimentally observed sharp enhancement of $\rho_{yx}$ in Phases II and IV cannot be reproduced by $\rho_{yx}^{\rm A}$, which suggests that these anomalies rather originate from the topological Hall term $\rho_{yx}^{\rm T}$.

The RXS experiments for GdRu$_2$Ge$_2$ have revealed that Phases II and IV are characterized by skyrmion lattice spin textures with skyrmion density $n_{\rm sk} \sim 1/(2.7$ nm$)^2$. The normal Hall coefficient $R_0$ (= 7.5 n$\Omega$ cm/T at 6 K) can be estimated from the slope of $\rho_{yx}$-$B$ profile in the saturated ferromagnetic state (Fig. 1j in the main text). If we assume the spin polarization ratio $P \sim 0.01$ (in the same order as $P \sim 0.07$ reported for Gd$_2$PdSi$_3$\cite{Gd2PdSi3}), the expected amplitude of topological Hall signal is $\rho_{yx}^{\rm T} = P R_0 (h/e) n_{\rm sk} \sim $ 0.04 $\mu\Omega$ cm. This well reproduces the observed amplitude of peak-like $\rho_{yx}$ anomalies for Phases II and IV in Supplementary Fig. 11c, indicating that they can be reasonably ascribed to the topological Hall effect associated with skyrmion spin texture.

Note that the large change of $\rho_{yx}$ value is also observed at the transition between the Phase V (square vortex lattice spin state with $N_{\rm sk} = 0$) and higher-$B$ phases (Figs. 1h and j in the main text). Similar behavior has also been reported for the isostructural GdRu$_2$Si$_2$\cite{GdRu2Si2}, although their microscopic origin is not clear at this stage. 
According to latest theoretical studies, the emergence of chiral Hall effect\cite{ChiralHall}, associated with the close link between the real- and momentum-space Berry phase, is often expected for non-collinear magnet even with $N_{\rm sk} = 0$. In addition, nontrivial Hall response may originate from various scattering processes such as electron-phonon, electron-skyrmion and skyrmion-skyrmion interactions\cite{NoncollinearHall, THE_JPhys, THE_PRB1, THE_PRB2}. To fully understand the $\rho_{yx}$-$B$ profiles for GdRu$_2$Ge$_2$, further theoretical studies and the detailed information on the electron, magnon and phonon band structures would be required.

\section{Future perspective}

In this section, we discuss several remaining issues for the future study. 

\subsection{Enhancement of magnetic ordering temperature.} Toward the potential application, further enhancement of magnetic ordering temperature is important. The present theoretical model based on Eq. (6) in the main text suggests that the rare-earth intermetallic compounds with highly symmetric crystal lattice would be promising for the realization of multi-step topological phase transitions. To enhance the magnetic ordering temperature, the employment of (a) higher density of magnetic rare-earth ions and/or (b) magnetic $3d$-transition metal ion would be effective. Previously, many room temperature rare-earth ferromagnets, such as SmCo$_5$\cite{SmCo5} and Gd$_7$Pd$_3$\cite{Gd7Pd3}, have been identified based on these approaches. By tuning the balance of RKKY interactions in such systems, the multi-step topological phase transition may be realized at higher temperature.

\subsection{Real-space observation of spin texture.} 

In principle, the Lorentz transmission electron microscopy (LTEM) technique will allow the real-space imaging of spin texture in each phase. Nevertheless, the magnetic modulation period in GdRu$_2$Ge$_2$ (2.7 nm) is almost close to the resolution limit of this technique, and such experiments are extremely difficult. In Ref. \cite{GdRu2Si2}, some of the present authors reported the atomic-resolution LTEM images in the square skyrmion lattice state, while this has been achieved by using the magnetic condenser-objective lens that inevitably generates 1.95 T of out-of-plane magnetic field. In other words, such a high spatial resolution cannot be achieved for different amplitude of magnetic field. We have once tried LTEM for GdRu$_2$Ge$_2$, while no meaningful magnetic contrast could be obtained due to the above reason. The direct real-space observation of spin texture is the issue for the future study.

\subsection{Skyrmions with higher density.} 

When we consider skyrmion particles as information carriers, further exploration of smaller size of skyrmions would be important to achieve the higher information density\cite{SkXReviewTokura}. According to the present theoretical model based on Eq. (6) in the main text, the skyrmion diameter is governed by the RKKY interaction and  electronic band structure. To realize smaller size of skyrmions, the tuning of Fermi-surface and associated nesting vector may be effective. In addition, the amplitude of fictitious magnetic field is proportional to the skyrmion density, and such a small size of skyrmion is expected to host a giant topological Hall effect. This phenomenon may contribute to the efficient electrical readout of topological magnetic states\cite{THE, Gd2PdSi3}.

\subsection{Response to various external stimuli.}

As discussed in Supplementary Note X, the conduction electrons interacting with skyrmion spin texture with non-zero topological charge $N_{\rm sk}$ feel a fictitious magnetic field in proportion to the skyrmion density, and it leads to the appearance of topological Hall effect\cite{THE}. In the similar manner, various off-diagonal response associated with fictitious magnetic field, such as topological magneto-optical Kerr effect\cite{TopologicalKerr} and topological Nernst effect\cite{TNE}, can be expected.

In principle, the observed magnetic quasi-particles (elliptic skyrmion, meron/anti-meron pair, and circular skyrmion) in GdRu$_2$Ge$_2$ will be transformable into each other by various external stimuli (such as electric current and mechanical strain), because of their pseudo degeneracy and sizable energy barrier\cite{Oike}. The further exploration of the above phenomena would be an interesting topic for future study.

\color{black}

\clearpage